\documentclass[12pt,letterpaper]{article}
\usepackage[utf8]{inputenc}
\usepackage{amsmath}
\usepackage{amsfonts}
\usepackage{amssymb}
\usepackage{amsthm}
\usepackage{bm}
\usepackage[table]{xcolor}
\usepackage{mathrsfs}
\usepackage{booktabs}
\usepackage[shortlabels]{enumitem}
\usepackage{graphicx}
\usepackage[left=2cm,right=2cm,top=2cm,bottom=2cm]{geometry}
\author{Matthew Durey${}^1$ and Paul A. Milewski${}^2$}
\title{Resonant triad interactions of gravity waves \\ in cylindrical basins}

\date{
\footnotesize
$^1$School of Mathematics and Statistics, University of Glasgow, University Place, Glasgow G12 8QQ, UK \\%
$^2$Department of Mathematical Sciences, University of Bath, Claverton Down, Bath BA2 7AY, UK
}

\begin{document}
\maketitle

\newcommand{\x}{\bm{x}}
\newcommand{\red}[1]{\textcolor{red}{#1}}
\newcommand{\sd}[2]{\frac{\mathrm{d} {#1}}{\mathrm{d} {#2}}}
\newcommand{\pd}[2]{\frac{\partial {#1}}{\partial {#2}}}
\newcommand{\DtN}{\mathscr{L}}
\newcommand{\DtNhat}{\hat{\mathscr{L}}}
\newcommand{\dA}{\,\mathrm{d}A}
\newcommand{\dr}{\,\mathrm{d}r}
\newcommand{\dth}{\,\mathrm{d}\theta}
\newcommand{\me}{\mathrm{e}}
\newcommand{\mi}{\mathrm{i}}
\newcommand{\cc}{\mathrm{c.c.}}
\newcommand{\Jb}{\mathrm{J}}
\newcommand{\Yb}{\mathrm{Y}}
\newcommand{\etal}{\emph{et al.}}

\newtheorem{theorem}{Theorem}
\newtheorem{lemma}{Lemma}

\definecolor{DarkGrey}{gray}{0.65}
\definecolor{Grey}{gray}{0.8}
\definecolor{LightGrey}{gray}{0.93}

\newcommand{\refone}[1]{\textcolor{black}{#1}}
\newcommand{\reftwo}[1]{\textcolor{black}{#1}}

\begin{abstract}
We present the results of a theoretical investigation into the existence, evolution and excitation of resonant triads of nonlinear free-surface gravity waves confined to a cylinder of finite depth. It is well known that resonant triads are impossible for gravity waves in laterally unbounded domains; we demonstrate, however, that horizontal confinement of the fluid may induce resonant triads for particular fluid depths.
For any three correlated wave modes arising in a cylinder of arbitrary cross-section, we prove necessary and sufficient conditions for the existence of a depth at which nonlinear resonance may arise, and show that the resultant critical depth is unique. We enumerate the low-frequency triads for circular cylinders, including a new class of resonances between standing and counter-propagating waves, and also briefly discuss annular and rectangular cylinders. Upon deriving the triad amplitude equations for a finite-depth cylinder of arbitrary cross-section, we deduce that the triad evolution is always periodic, and determine parameters controlling the efficiency of energy exchange. In order to excite a particular triad, we explore the influence of external forcing; in this case, the triad evolution may be periodic, quasi-periodic, or chaotic. Finally, our results have potential implications on resonant water waves in man-made and natural basins, such as industrial-scale fluid tanks, harbours and bays.
\end{abstract}

\section{Introduction}
\label{sec:intro}

Nonlinear resonance is a mechanism by which energy is continuously transferred between a small number of linear wave modes. This phenomenon, first observed in Wilton's analysis of gravity-capillary wave trains \cite{Wilton1915}, has been the subject of frequent investigation over the past century \cite{McGoldrick1965, McGoldrick1970b, McGoldrick1970a, Simmons1969, SchwartzVandenBroeck1979, HammackHenderson1993, CraikBook}; indeed, nonlinear resonance has since been observed for wave trains in a growing number of dispersive wave systems, including 
gravity waves \cite{Phillips1960, Hasselmann1961, LonguetHiggins1962, Benney1962}, 
acoustic-gravity waves \cite{Kadri2013, Kadri2016}, 
flexural-gravity waves \cite{Wang2013},
two-layer flows \cite{Ball1964, Joyce1974, Segur1980}, and 
atmospheric flows \cite{Raupp2008, Raupp2009}. Whilst the aforementioned studies typically consider nonlinear resonance for laterally unbounded domains, the purpose of this study is to demonstrate that energy exchange between free-surface gravity waves may be induced and accentuated by horizontal confinement.

We focus our study on the collective resonance of three linear wave modes, henceforth referred to as a \emph{triad} \cite{Bretherton1964}. In laterally unbounded domains, the monotonic and concave form of the dispersion curve precludes the existence of resonant triads for gravity wave trains at finite depth \cite{Phillips1960, Hasselmann1961}, with resonant quartets instead being the smallest possible collective resonant interaction \cite{Benney1962, LonguetHiggins1962, BergerMilewski2003}. However, confinement of the fluid to a vertical cylinder results in linear wave modes that differ in form to sinusoidal plane waves (except for a rectangular cylinder), so the preclusion of resonant triads no longer applies.
Indeed, our study demonstrates that, under certain conditions, resonant triads may arise in cylinders of arbitrary cross-section for specific values of the fluid depth. As resonant triads evolve over a much faster time scale than that of resonant quartets, the exchange of energy in gravity waves is thus more efficient under the influence of lateral confinement \cite{Michel2019}, with potential implications on resonant sloshing in man-made and natural basins \cite{Bryant1989}.

Prior investigations of confined resonant free-surface gravity waves have predominantly focused on the so-called 1:2 resonance, which arises when two of the three linear wave modes comprising a triad coincide.
For axisymmetric standing waves in a circular cylinder,
Mack \cite{Mack1962} determined a condition for the existence of critical depth-to-radius ratios at which a 1:2 resonance may arise, a result later generalised to cylinders of arbitrary cross-section \cite{Miles1984b}.
Miles \cite{Miles1976, Miles1984a} then characterised the weakly nonlinear evolution of such internal resonances, demonstrating that a 1:2 resonance is impossible in a rectangular cylinder \cite{Miles1976}. Although Miles' seminal results provide an informative view of the weakly nonlinear dynamics, the influence of fully nonlinear effects was later assessed by Bryant \cite{Bryant1989} and Yang \emph{et al.}\ \cite{Yang2021}. For the case of a circular cylinder of finite depth, Bryant \cite{Bryant1989} and Yang \emph{et al.}\ \cite{Yang2021} characterised new steadily propagating nonlinear waves arising in the vicinity of a 1:2 resonance, and Yang \emph{et al.}\ \cite{Yang2021} also computed nonlinear near-resonant axisymmetric standing waves.
Finally, broader mathematical properties of water waves exhibiting O(2) symmetry (of which a circular cylinder is one example) were analysed by Bridges \& Dias \cite{BridgesDias1990} and Chossat \& Dias \cite{ChossatDias1995}.

Given the restrictive set of critical depths at which a 1:2 resonance may arise \cite{Bryant1989,Yang2021}, it is natural to explore the possibility of nonlinear resonance in cylinders whose depth departs from the depths that trigger a 1:2 resonance.
To the best of our knowledge, the first and only such study was the seminal experimental investigation performed by Michel \cite{Michel2019}, who focused on resonant triads arising for free-surface gravity waves confined to a finite-depth circular cylinder. Notably, the cylinder depth in Michel's experiment was judiciously chosen so as to isolate a specific triad. Michel utilised bandlimited random horizontal vibration so as to excite two members of the triad, whose nonlinear interaction led to the growth of the third mode. Significantly, the energy of the third mode was, on average, the product of the energies of the remaining two modes, thereby satisfying the quadratic energy exchange typical of resonant triads.

In order to exemplify the mechanism of nonlinear resonance, Michel \cite{Michel2019} also calculated the response of a child mode due to the nonlinear interaction between two parent modes (where all three wave modes comprise the triad).
Notably, Michel's calculation is restricted to the early stages of growth and to particular relative phases of the wave modes. In addition, Michel considered a fluid of infinite depth for all but the resonance conditions, for which finite-depth corrections were included. In contrast, we consider general resonances in arbitrary cylinders of finite depth and derive equations for the triad evolution over long time-scales. We also believe some nonlinear contributions to the interactions were omitted from Michel's calculation, resulting in quantitative differences (see \S \ref{sec:multiple_scales_summary}).

The goal of our study is to unify the existence and evolution of 1:2 and triadic resonances into a single mathematical framework, effectively characterising all triad interactions of this type.
Based on existing theory, it is unclear how the existence of resonant triads depends on the form of the cylinder cross-section, and which combinations of wave modes are permissible for judicious choice of the fluid depth. Furthermore, the range of depths that may excite a particular triad is uncertain, with 1:2 resonances only excited in a very narrow window about each critical depth \cite{Mack1962, Miles1984b}. Once a particular triad is excited, one anticipates that the triad evolution will be governed by the canonical triad equations \cite{Bretherton1964, CraikBook}; however, quantifying the triad evolution and relative energy exchange requires computation of the triad coupling coefficients. Finally, it is unclear how best to excite triads in arbitrary cylinders, both with and without external forcing.

We here present a relatively comprehensive characterisation of the existence, evolution and excitation of resonant triads for gravity waves confined to a cylinder of arbitrary cross-section and finite depth. In order to reduce the problem to its key components, we first truncate the Euler equations, recasting the fluid evolution in terms of a finite-depth Benney-Luke equation (\S \ref{sec:formulation}), incorporating only the nonlinear interactions necessary for resonant triads. In \S \ref{sec:triads_existence}, we prove necessary and sufficient conditions for there to exist a finite depth at which three linear wave modes may form a resonant triad. In particular, we prove that resonant triads are impossible for rectangular cylinders, yet there is an abundance of resonant triads for circular cylinders. We then use multiple-scales analysis to determine the long-time evolution of a triad in a cylinder of arbitrary cross-section (\S \ref{sec:triad_eqs}), from which we characterise the relative coupling of different triads. Finally, we explore the excitation of resonant triads (\S \ref{sec:excitation}), and discuss the potential extension of our theoretical developments to the cases of applied forcing and two-layer flows (\S \ref{sec:discussion}).

\section{Formulation}
\label{sec:formulation}

\begin{figure}
\centering
\includegraphics[width=0.25\textwidth]{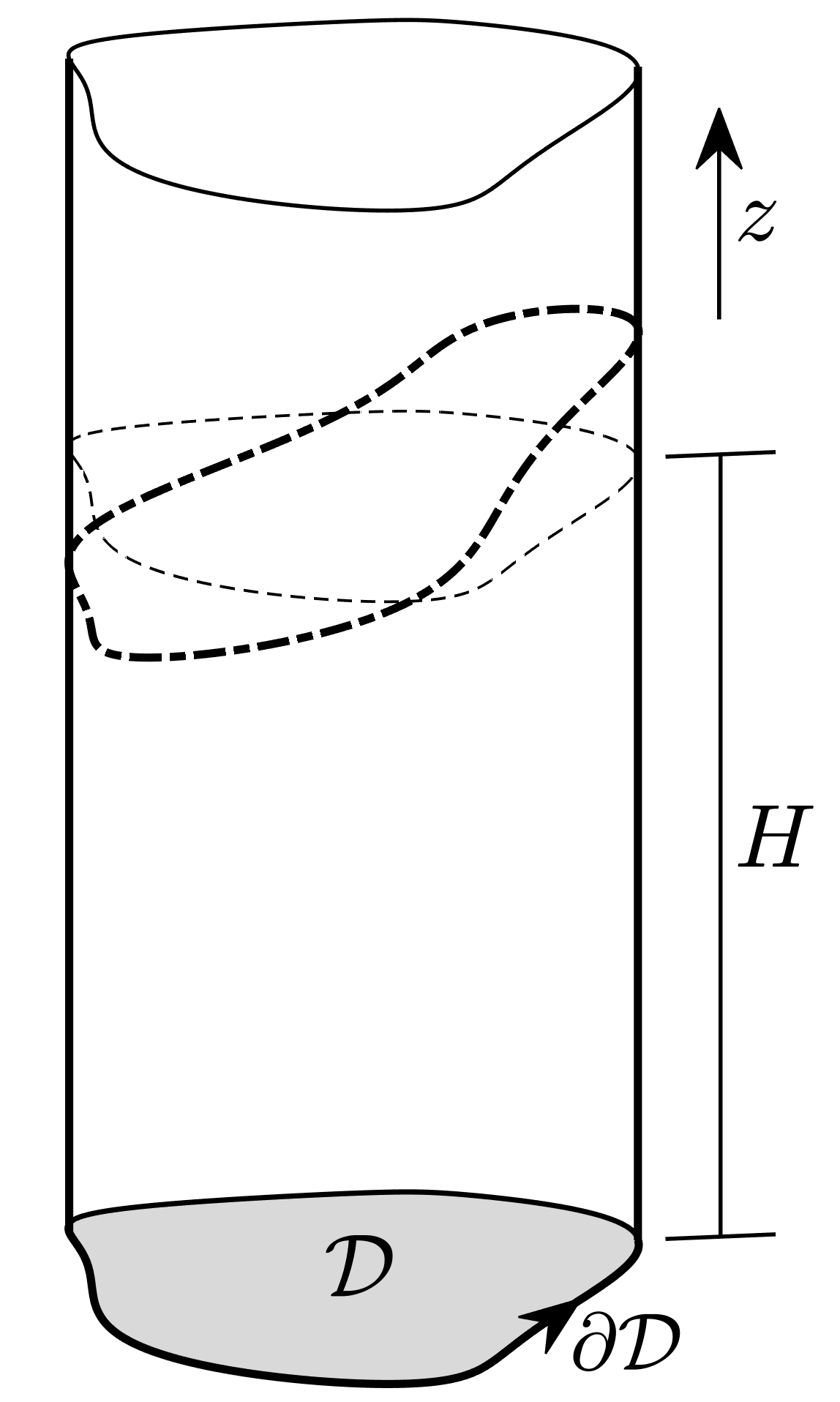}
\caption{\label{fig:Schematic_diagram} Schematic diagram of the cylindrical tank (with cross-section $\mathcal{D}$ and boundary $\partial \mathcal{D}$) partially filled with liquid. The undisturbed free surface (dashed lines) lies on $z = 0$, a distance $H$ above the rigid bottom plane (grey). The disturbed free surface is sketched in dash-dotted lines.}
\end{figure}

We consider the irrotational flow of an inviscid, incompressible liquid that is bounded above by a free surface, confined laterally by the vertical walls of a cylinder whose horizontal cross-section, $\mathcal{D}$, is enclosed by the curve $\partial\mathcal{D}$, and bounded below by a rigid horizontal plane lying a distance $H$ below the undisturbed free surface; see figure \ref{fig:Schematic_diagram}. We consider the fluid evolution in dimensionless variables, taking the cylinder's typical horizontal extent, $a$, as the unit of length, and $\sqrt{ag^{-1}}$ as the unit of time, where $g$ is the acceleration due to gravity. It follows that the dimensionless free-surface elevation, $\eta(\x,t)$, and velocity potential, $\phi(\x,z,t)$, evolve according to the equations
\begin{subequations}
\label{eq:Euler}
\begin{alignat}{2}
\Delta \phi + \phi_{zz} &= 0 \qquad && \mathrm{for}\,\,\,\x \in \mathcal{D}, \quad -h < z < \epsilon \eta, \label{eq:Euler_Laplace} \\
\phi_t + \eta + \frac{\epsilon}{2}\Big(|\nabla \phi|^2 + \phi_z^2\Big) &= 0 \quad &&\mathrm{for}\,\,\,\x \in \mathcal{D}, \quad z = \epsilon \eta, \label{eq:Euler_DBC} \\
\eta_t + \epsilon \nabla\phi \cdot \nabla\eta &= \phi_z \quad  &&\mathrm{for}\,\,\,\x \in \mathcal{D}, \quad z = \epsilon \eta, \label{eq:Euler_KBC} \\
\bm{n}\cdot \nabla \phi &= 0 &&\mathrm{for}\,\,\,\x \in \partial\mathcal{D}, \quad -h < z < \epsilon \eta, \label{eq:Euler_no_flux_walls} \\
\phi_z &= 0 &&\mathrm{for}\,\,\,\x \in \mathcal{D}, \quad z = -h, \label{eq:Euler_no_flux_base} 
\end{alignat}
\end{subequations} 
corresponding to the continuity equation, dynamic and kinematic boundary conditions, and no-flux through the vertical walls and horizontal base, respectively.
In equation \eqref{eq:Euler}, the dimensionless parameter $\epsilon$ is proportional to the typical wave slope, $h = H/a$ is the ratio of the fluid depth to the typical horizontal extent, $\bm{n}$ is a unit vector normal to the boundary $\partial \mathcal{D}$, and the operators $\nabla$ and $\Delta$ denote the horizontal gradient and Laplacian, respectively. Moreover, conservation of mass implies that the free surface satisfies $\iint_{\mathcal{D}} \eta \dA = 0$ for all time. Finally, in dimensional variables, $a\x$ is the two-dimensional horizontal coordinate, $az$ is the upward-pointing vertical coordinate, $\sqrt{ag^{-1}} t$ denotes time, $\epsilon a \eta$ is the free-surface displacement, and $\epsilon a \sqrt{ag}\phi$ is the velocity potential.

We aim to develop a broad framework for understanding resonant triads in a cylinder of finite depth; however, care must be taken when modelling fluid-boundary interactions and determining the class of permissible cylinder cross-sections.  
From a modelling perspective, we employ an assumption generally implicit to the water-wave problem in bounded domains; specifically, we neglect the meniscus and dissipation arising near the vertical walls \cite{Miles1967}, thus determining that the free surface intersects the boundary normally, i.e.\ $\bm{n}\cdot \nabla \eta = 0$ for $\x \in \partial\mathcal{D}$ \cite{MilesHenderson1990}.
In order to maximise the generality of our investigation, we allow the cylinder cross-section, $\mathcal{D}$, to be fairly arbitrary; however, the mathematical developments presented herein require $\mathcal{D}$ to be bounded with a piecewise-smooth boundary, thereby allowing us to utilise the spectral theorem for compact self-adjoint operators \cite{KreyszigBook} and the divergence theorem. As most cylinders of practical interest consist of a piecewise-smooth boundary, this mathematical restriction fails to limit the breadth of our study.

\subsection{Derivation of the Benney-Luke equation}
\label{sec:BL_eq}

As our study is focused on the weakly nonlinear evolution of small-amplitude waves, we proceed to simplify \eqref{eq:Euler} in the case $0 < \epsilon \ll 1$ and $h = O(1)$. We begin by expanding the dynamic and kinematic boundary conditions (equations \eqref{eq:Euler_DBC}--\eqref{eq:Euler_KBC}) about $z = 0$ in powers of $\epsilon$, which, upon eliminating $\eta$, gives rise to the equation \cite{Benney1962, MilewskiKeller1996}
\begin{equation}
\label{eq:BL_intermediate}
\phi_{tt} + \phi_z = \epsilon\Big(\partial_t (\phi_{tz}\phi_t) + \phi_{zz}\phi_t - \partial_t(|\nabla \phi|^2) - \phi_z\phi_{zt}\Big) + O(\epsilon^2) \quad\mathrm{for}\quad \x \in \mathcal{D}, \quad z = 0.
\end{equation} 
To reduce the fluid evolution to the dynamics arising on the linearised free surface, $z = 0$, we define the Dirichlet-to-Neumann operator, $\DtN$, so that $\DtN \phi|_{z = 0} = \phi_z|_{z = 0}$. Here $\phi$ satisfies Laplace's equation \eqref{eq:Euler_Laplace} over the linearised domain $-h < z < 0$, with $\partial_z \phi = 0$ on $z = -h$ (see equation \eqref{eq:Euler_no_flux_base}) and $\bm{n}\cdot \nabla \phi = 0$ for $\x \in \partial \mathcal{D}$ (see equation \eqref{eq:Euler_no_flux_walls}). Notably, the Dirichlet-to-Neumann operator may be defined in terms of its spectral representation, as detailed in \S \ref{sec:DtN_operator}. By denoting  $u(\x,t) = \phi(\x,0,t)$, we finally obtain the finite-depth Benney-Luke equation \cite{Benney1962, BenneyLuke1964, MilewskiKeller1996}
\begin{equation}
\label{eq:BL_eq}
u_{tt} + \DtN u + \epsilon\bigg(
u_t\big(\DtN^2 + \Delta\big)u + \pd{}{t}\Big[(\DtN u)^2 + |\nabla u|^2\Big]\bigg) =  O(\epsilon^2) \quad\mathrm{for}\quad \x \in \mathcal{D},
\end{equation} 
where we have simplified the nonlinear terms in equation \eqref{eq:BL_intermediate} using $\phi_{zz} = -\Delta \phi$ and  $u_{tt} = -\DtN u + O(\epsilon)$. 

The remainder of our investigation will be focused on the evolution of resonant triads governed by the Benney-Luke equation \eqref{eq:BL_eq}.
As resonant triads arising in confined geometries are governed primarily by quadratic nonlinearities, it is sufficient to neglect terms of size $O(\epsilon^2)$ in equation \eqref{eq:BL_eq}; however, higher-order corrections to the Benney-Luke equation may be derived by following a similar expansion procedure \cite{Benney1962, MilewskiKeller1996, BergerMilewski2003}.
Although our investigation is mainly focused on the evolution of the velocity potential, $u$, one may recover the leading-order free-surface elevation from the dynamic boundary condition \eqref{eq:Euler_DBC}, namely $\eta = -u_t + O(\epsilon)$.

\subsection{Spectral representation of the Dirichlet-to-Neumann operator}
\label{sec:DtN_operator}

The Dirichlet-to-Neumann operator, $\DtN$, may be understood in terms of the discrete set of orthogonal eigenfunctions of the horizontal Laplacian operator \cite{KreyszigBook}. Specifically, we consider the set of real-valued eigenfunctions, $\Phi_n(\x)$, satisfying
\[ -\Delta \Phi_n = k_n^2 \Phi_n \,\,\, \mathrm{for}\,\,\, n = 0, 1, \ldots,\]
where the corresponding eigenvalues, $k_n^2$, are ordered so that $0 = k_0 < k_1\leq k_2 \leq \ldots$. Moreover, each eigenfunction satisfies the boundary condition $\bm{n}\cdot\nabla \Phi_n = 0$ on $\partial\mathcal{D}$, as motivated by the no-flux condition \eqref{eq:Euler_no_flux_walls}. Finally, the orthogonal eigenfunctions are normalised so that $\langle \Phi_m, \Phi_n \rangle = \delta_{mn}$, where
\[\langle f, g\rangle = \frac{1}{S}\iint_\mathcal{D} f g \dA \]
defines an inner product for real functions $f$ and $g$, $S$ is the area of $\mathcal{D}$, and $\delta_{mn}$ is the Kronecker delta. Notably, $\Phi_0(\x) = 1$ is the constant eigenfunction, with corresponding eigenvalue $k_0 = 0$.

To determine the Dirichlet-to-Neumann operator for sufficiently smooth $\phi$, we first substitute the series expansion $\phi(\x,z) = \sum_{n = 0}^\infty \phi_n(z)\Phi_n(\x)$ into Laplace's equation \eqref{eq:Euler_Laplace}, where we have temporally omitted the time dependence. We then solve the resulting equation for $\phi_n(z)$ over the linearised domain $-h < z < 0$, in conjunction with the no-flux condition on $z = -h$ (see equation \eqref{eq:Euler_no_flux_base}). It follows that
$\partial_z\phi_n(0) = \hat{\DtN}_n \phi_n(0)$, where
\begin{equation}
\label{eq:DtN_symbol}
\hat{\DtN}_n = k_n \tanh(k_n h)
\end{equation}
is the spectral multiplier of the Dirichlet-to-Neumann operator, $\DtN$. By expressing the time-dependent free-surface velocity potential, $u = \phi|_{z = 0}$, in terms of the basis expansion $u(\x, t ) = \sum_{n = 0}^\infty u_n(t) \Phi_n(\x)$, it follows that the Dirichlet-to-Neumann map has the spectral representation $\DtN u = \sum_{n = 0}^\infty \hat{\DtN}_n u_n \Phi_n$.

\section{The existence of resonant triads}
\label{sec:triads_existence}

Resonant triads arise due to the exchange of energy between linear wave modes, an effect induced by nonlinear wave interactions. In order to define resonant triads mathematically, it is necessary to first determine the angular frequency associated with each linear wave mode. In the limit $\epsilon \rightarrow 0$, the Benney-Luke equation \eqref{eq:BL_eq} reduces to the linear equation $u_{tt} + \DtN u = 0$. By seeking a solution to the linearised Benney-Luke equation of the form $u(\x,t) = \Phi_n(\x) \me^{-\mi \omega_n t}$, we conclude that the angular frequency, $\omega_n$, satisfies $\omega_n^2 = \hat{\DtN}_n$, or the more familiar \cite{LambBook}
\begin{equation}
\label{eq:dis_relation}
\omega_n^2 = k_n \tanh(k_n h).
\end{equation}
 As we will see, a crucial aspect of the following analysis is that the angular frequency depends on the fluid depth, i.e.\ $\omega_n(h)$. 
Finally, we note that the angular frequency is larger for more oscillatory eigenfunctions (i.e.\ for larger values of $k_n$); by analogy to the evolution of plane gravity waves, we refer to $k_n$ as a `wavenumber' henceforth.

We proceed by considering three linear wave modes, enumerated $n_1$, $n_2$ and $n_3$, where we denote
\[\Omega_j = \omega_{n_j}, \quad K_j = k_{n_j}, \quad \mathrm{and}\quad \Psi_j(\x) = \Phi_{n_j}(\x) \quad \mathrm{for}\,\,\, j = 1, 2, 3.  \]
Notably, we exclude the wavenumber $k_0 = 0$ from consideration as the corresponding eigenmode, $\Phi_0$, simply reflects the invariance of the Benney-Luke equation \eqref{eq:BL_eq} under the mapping $u \mapsto u + \mathrm{constant}$; henceforth, we consider only wavenumbers $K_j > 0$. The three linear wave modes form a resonant triad if there is a critical fluid depth, $h_c$, satisfying
\begin{equation}
\label{eq:triad_sum_gen}
\Omega_1(h_c) \pm \Omega_2(h_c) \pm \Omega_3(h_c) = 0,
\end{equation}
where all four sign combinations are permissible (we consider $\Omega_j > 0$ without loss of generality). To simplify notation in the following arguments, we restrict our attention to the particular case
\begin{equation}
\label{eq:triad_sum1}
\Omega_1(h_c) + \Omega_2(h_c) = \Omega_3(h_c),
\end{equation}
where the other three sign combinations in equation \eqref{eq:triad_sum_gen} may be recovered by suitable re-indexing of the $\Omega_j$ terms.
However, as we will see in \S \ref{sec:triad_eqs}, an additional constraint necessary for triads to exist is the eigenmode correlation condition, 
\begin{equation}
\label{eq:corr_cond}
\iint_{\mathcal{D}} \Psi_1\Psi_2\Psi_3 \dA  \neq 0,
\end{equation}
which implies that the product of any two eigenmodes is non-orthogonal to the remaining eigenmode.

\subsection{The existence of a critical depth}
\label{sec:critical_depth_existence}

We proceed to determine necessary and sufficient conditions on the wavenumbers, $K_j$, for there to exist a depth, $h_c$, at which a resonant triad forms, where such a critical depth is unique. We summarise our results in terms of the following theorem.

\begin{theorem}
\label{thm:triads}
There exists a positive and finite value of $h$ such that $\Omega_1 + \Omega_2 = \Omega_3$ if and only if
\begin{equation}
\label{eq:kineq}
K_1 + K_2 < K_3 < \big(\sqrt{K_1} + \sqrt{K_2}\big)^2. 
\end{equation}
When this pair of inequalities is satisfied, the corresponding value of $h$ is unique.
\end{theorem}
We briefly sketch the proof of Theorem \ref{thm:triads}, with full details presented in appendix \ref{app:thm_proof}. 
We first demonstrate that no solutions to $\Omega_1 + \Omega_2 = \Omega_3$ are possible when the bounds in equation \eqref{eq:kineq} are violated, i.e.\ when $K_1 + K_2 \geq K_3$ or when $\sqrt{K_1} + \sqrt{K_2} \leq \sqrt{K_3}$. We then consider the case where the inequalities \eqref{eq:kineq} are satisfied and determine the existence of positive roots to the function $F(h) = (\Omega_1(h) +\Omega_2(h))/\Omega_3(h) - 1$. In this case, we demonstrate that $\lim_{h \rightarrow 0} F(h) < 0$ and $\lim_{h\rightarrow\infty} F(h) > 0$, from which we conclude that $F(h)$ has at least one root (by continuity of $F$). Finally, we deduce that this root is unique by proving that $F(h)$ is a strictly monotonically increasing function of $h$ when the inequalities \eqref{eq:kineq} are satisfied.

Two important conclusions may be deduced from Theorem \ref{thm:triads}.
First, it follows from equation \eqref{eq:kineq} that the wavenumber, $K_3$, corresponding to the largest angular frequency, $\Omega_3$, is larger than both the other two wavenumbers ($K_1$ and $K_2$), but it cannot be arbitrarily large (as supplied by the upper bound). For a given pair of eigenmodes (say $\Psi_1$ and $\Psi_2$), we conclude that there are likely to be only finitely many eigenmodes that can resonate with this pair (indeed, that number might fairly small, or even zero).
Second, when modes 1 and 2 coincide (a 1:2 resonance), one deduces that $\Omega_1 = \Omega_2$ and $K_1 = K_2$; as such, the existence bounds \eqref{eq:kineq} simplify to $2K_1 < K_3 < 4K_1$, or $2 < K_3/K_1 < 4$ \cite{Mack1962, Miles1984b}.

\subsection{Determining the critical depth}
\label{sec:critical_depth_determine}

Although Theorem \ref{thm:triads} determines necessary and sufficient conditions on the wavenumbers, $K_j$, for there to be a critical depth, $h_c$, at which a resonant triad exists, the critical depth remains to be determined. In general, the critical depth must be computed numerically (being the unique root of the nonlinear function $F(h)$); however, we demonstrate that useful quantitative and qualitative information may be obtained via asymptotic analysis. For the remainder of this section, we consider the rescaled wavenumbers, $\xi_1 = K_1/K_3$ and $\xi_2 = K_2/K_3$, and the rescaled depth, $\zeta = K_3 h$; it remains to determine the root, $\zeta_c$, of 
\begin{equation}
\label{eq:resc_freq_sum}
F(\zeta) = \sqrt{\frac{ \xi_1 \tanh(\xi_1 \zeta )} {\tanh(\zeta)} } + \sqrt{\frac{ \xi_2 \tanh(\xi_2 \zeta )} {\tanh(\zeta)} } - 1
\end{equation}
when $\xi_1, \xi_2 > 0$ satisfy
\begin{equation}
\label{eq:xi_ineq}
\xi_1 + \xi_2 < 1 < \sqrt{\xi_1} + \sqrt{\xi_2}.
\end{equation}
\begin{figure}
\centering
\includegraphics[width=0.95\textwidth]{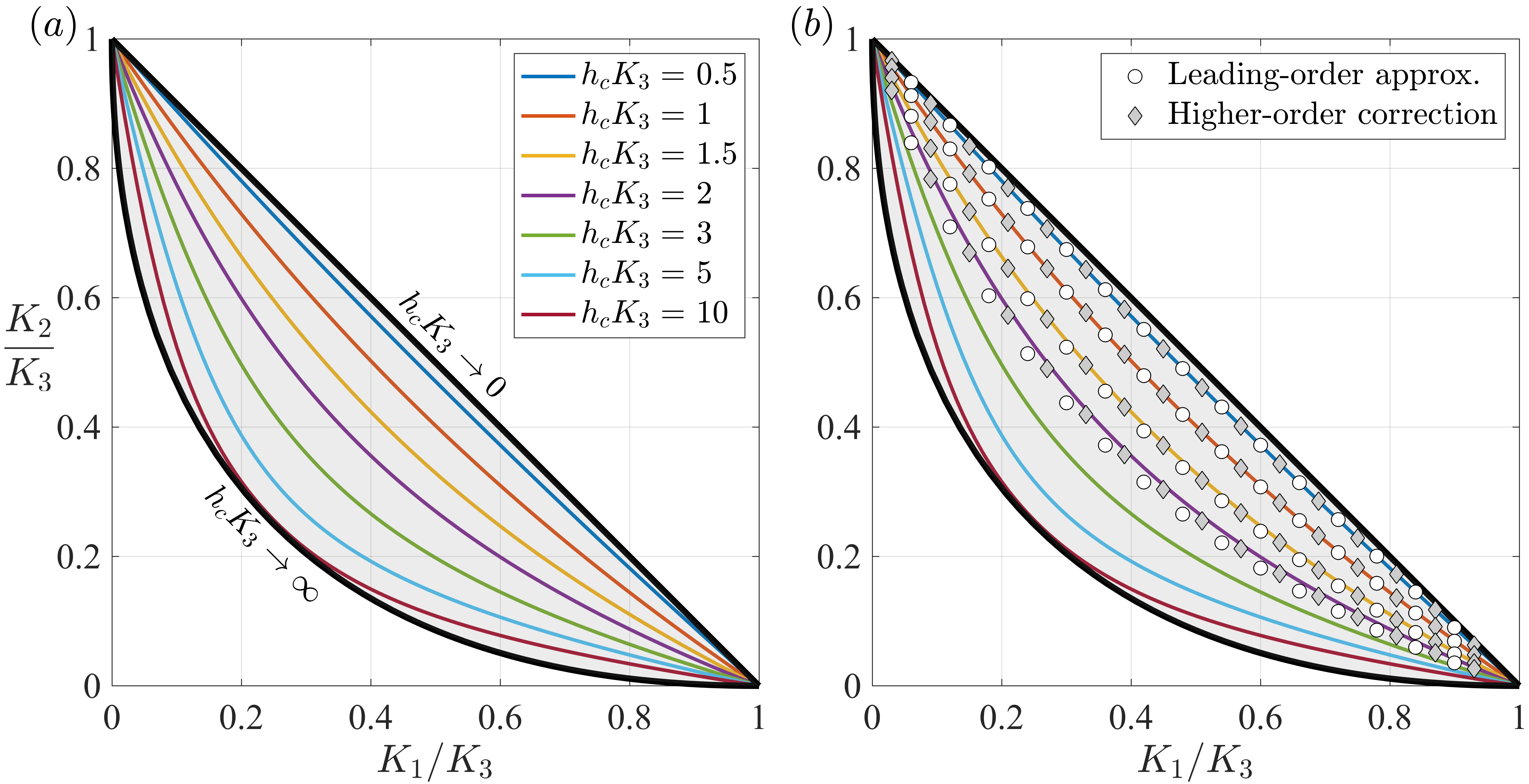}
\caption{\label{fig:Contours_of_h} Contours of the rescaled critical depth, $\zeta_c = h_cK_3$, as a function of the rescaled wavenumbers, $\xi_1 = K_1/K_3$ and $\xi_2 = K_2/K_3$.
$(a)$ The contours computed numerically from equation \eqref{eq:resc_freq_sum}. The black lines indicate the limiting cases of $\zeta_c \rightarrow 0$ (at $\xi_1 + \xi_2 = 1$) and $\zeta_c \rightarrow \infty$ (at $\sqrt{\xi_1} + \sqrt{\xi_2} = 1$).
$(b)$ The contours are overlaid by the leading-order approximation (equation \eqref{eq:zeta*_cont1}; circles) and the higher-order correction (equation \eqref{eq:zeta*_exp_higher_order}; diamonds) for $\zeta_c$ equal to 0.5, 1, 1.5 and 2.}
\end{figure}
In figure \ref{fig:Contours_of_h}$(a)$, we present contours of the critical rescaled depth, $\zeta_c$, in the $(\xi_1, \xi_2)$-plane, restricted to the region demarcated by equation \eqref{eq:xi_ineq}. Consistent with the limits $\lim_{\zeta\rightarrow 0}F(\zeta) = \xi_1 + \xi_2 - 1$ and $\lim_{\zeta\rightarrow \infty}F(\zeta) = \sqrt{\xi_1} + \sqrt{\xi_2} - 1$, we observe that the root, $\zeta_c$, tends to zero at the line $\xi_1 + \xi_2 = 1$, and approaches infinity at the curve $\sqrt{\xi_1} + \sqrt{\xi_2} = 1$. Furthermore, the uniqueness of the root of $F$ for given $(\xi_1, \xi_2)$ is reflected in the observation that the contours of $\zeta_c$ do not cross. Finally, we note that the contours are symmetric about the line $\xi_1 = \xi_2$, which is a direct consequence of the invariance of $F(\zeta)$ under the mapping $\xi_1 \leftrightarrow \xi_2$ (see equation \eqref{eq:resc_freq_sum}).

Although we are primarily interested in the physically relevant case for which the cylinder's depth-to-width ratio, $h$, is of size $O(1)$, an informative analytic result may be obtained by considering $F(\zeta)$ in the limit $\zeta \ll 1$ (or $K_3 h \ll 1$). By utilising the Taylor expansion
$$\sqrt{\tanh(x)} \sim \sqrt{x}\bigg(1 - \frac{x^2}{6} + \frac{19}{360}x^4 + O\big(x^6\big)\bigg),$$
we obtain
\begin{equation}
\label{eq:om_exp}
\sqrt{\xi_1\tanh(\xi_1\zeta)} + \sqrt{\xi_2\tanh(\xi_2\zeta)} - \sqrt{\tanh(\zeta)} \sim \sqrt{\zeta}\bigg[\Big(\xi_1 + \xi_2 - 1\Big) - \frac{\zeta^2}{6}\Big(\xi_1^3 + \xi_2^3 - 1\Big) + O(\zeta^4)\bigg]
\end{equation}
for $0 < \zeta \ll 1$. Whilst deriving equation \eqref{eq:om_exp}, we have utilised the bound $\xi_1, \xi_2 < 1$ (see equation \eqref{eq:xi_ineq}), which additionally ensures that $0 < \xi_j\zeta \ll 1$ for $j = 1,2$. We note that the left-hand side of equation \eqref{eq:om_exp} is equal to $F(\zeta)\tanh(\zeta)$, so $\zeta_c$ satisfies
\begin{equation}
\label{eq:zeta*_exp}
\xi_1 + \xi_2 - 1 - \frac{\zeta_c^2}{6}\Big(\xi_1^3 + \xi_2^3 - 1\Big) = O(\zeta_c^4),
\end{equation}
provided that $0 < \zeta_c \ll 1$. By neglecting terms of size $O(\zeta_c^4)$ in equation \eqref{eq:zeta*_exp}, one may then easily solve for $\zeta_c$ in terms of $\xi_1$ and $\xi_2$.

Alternatively, a more succinct expression for $\zeta_c$ may be found by first noting that 
\begin{equation}
\label{eq:xi_cubed_exp}
\xi_1^3 + \xi_2^3 = (\xi_1 + \xi_2)^3 - 3\xi_1\xi_2(\xi_1 + \xi_2) = 1 - 3\xi_1\xi_2 + O(\zeta_c^2),
\end{equation}
where we have utilised the leading-order approximation $\xi_1 + \xi_2 = 1 + O(\zeta_c^2)$ (see equation \eqref{eq:zeta*_exp}) to determine the second equality.
Upon substituting equation \eqref{eq:xi_cubed_exp} into equation \eqref{eq:zeta*_exp}, we find that $\xi_1$, $\xi_2$ and $\zeta_c$ are now related by the notably simpler expression
\begin{equation}
\label{eq:zeta*_exp_simp}
\xi_1 + \xi_2 - 1 + \frac{\zeta_c^2}{2}\xi_1\xi_2 = O(\zeta_c^4).
\end{equation}
By neglecting terms of $O(\zeta_c^4)$, the leading-order approximation for the rescaled critical depth, $\zeta_c$, is given by
\begin{equation}
\label{eq:zeta*_quad}
\zeta_c \sim \sqrt{\frac{2(1 - \xi_1 - \xi_2)}{\xi_1\xi_2}},
\end{equation}
an expression valid when $0 < \zeta_c \ll 1$ and $\xi_1 + \xi_2 < 1$ (see equation \eqref{eq:xi_ineq}). Alternatively, one may deduce from equation \eqref{eq:zeta*_exp_simp} that the contours of $\zeta_c$ satisfy the approximate form
\begin{equation}
\label{eq:zeta*_cont1}
\xi_2 \sim \frac{1 - \xi_1}{1 + \frac{1}{2}\zeta_c^2\xi_1},
\end{equation}
where the term in the denominator is responsible for the increased `bending' of the contours as $\zeta_c$ becomes progressively larger (see figure \ref{fig:Contours_of_h}). We note that the additional simplification afforded by equation \eqref{eq:xi_cubed_exp} allows for a far more tractable representation of the contours relative to solving equation \eqref{eq:zeta*_exp} directly for $\xi_2$ given $\xi_1$ and $\zeta_c$.

Despite being derived under the assumption $0 < \zeta_c \ll 1$, we see in figure \ref{fig:Contours_of_h}$(b)$ that the contours given by equation \eqref{eq:zeta*_cont1} agree favorably with the numerical solution even up to $\zeta_c \approx 1$. However, it is readily verified from equation \eqref{eq:zeta*_quad} that the asymptotic approximation of each contour crosses the boundary curve $\sqrt{\xi_1} + \sqrt{\xi_2} = 1$ at $\zeta_c = 4$ (for which $\xi_1 = \xi_2 = \frac{1}{4}$), thereby demonstrating that the reduced asymptotic form has limited applicability (even in a qualitative sense) for slightly larger values of $\zeta_c$. One may further improve the quantitative (and, to an extent, qualitative) agreement between the asymptotic analysis and numerical computation by including terms of size $O(\zeta^4)$ in equation \eqref{eq:om_exp}; indeed, an analogous calculation gives rise to the following higher-order correction to equation \eqref{eq:zeta*_exp_simp}:
\begin{equation}
\label{eq:zeta*_exp_higher_order}
\xi_1 + \xi_2 - 1 + \frac{\zeta_c^2}{2}\xi_1\xi_2 + \frac{\zeta_c^4}{72} \xi_1\xi_2\big(\xi_1\xi_2 - 1\big) = O(\zeta_c^6).
\end{equation}
Although one may then solve for $\zeta_c$ given $\xi_1$ and $\xi_2$ (or, alternatively, determine the contours of $\zeta_c$) by truncating terms of $O(\zeta_c^6)$ in equation \eqref{eq:zeta*_exp_higher_order}, the resulting algebraic expressions yield little qualitative information. However, one may, in principle, use this reduced form as a reasonable initial guess for a numerical root-finding algorithm for determining the root of $F(\zeta)$, provided that $\zeta_c$ is not too large.

\subsection{Example cavities}
\label{sec:example_cavities}

Our investigation into the emergence of resonant triads has been focused, thus far, on finite-depth cylinders with arbitrary horizontal cross-section. However, it is convenient to understand how the results of Theorem \ref{thm:triads} influence the formation (or not) of resonant triads for some specific cross-sections, namely rectangular, circular, and annular cylinders.

\subsubsection{Rectangular cylinder}
\label{sec:rectangular_cylinder}

It is well known that resonant triads are impossible for plane gravity waves evolving across an unbounded horizontal domain of finite depth \cite{Phillips1960, Hasselmann1961}.
\footnote{Weak interactions are possible, however, in the shallow-water limit, $K_jh\rightarrow 0$, for which $\tanh(K_jh)$ in the dispersion relation \eqref{eq:dis_relation} is replaced by its leading-order approximation, $K_j h$ \cite{Phillips1960, Bryant1973, Miles1976}.} 
We now utilise Theorem \ref{thm:triads} to demonstrate a similar result: resonant triads are impossible for gravity waves evolving within a rectangular cylinder of finite depth. Our result generalises the special case of a 1:2 resonance, for which the impossibility of internal resonance in a rectangular cylinder was demonstrated by Miles \cite{Miles1976}.

To proceed, we consider a rectangular cylinder with side lengths $L_x$ and $L_y$.
By orientating the Cartesian coordinate system, $\x = (x,y)$, so that the cylinder cross-section is defined by the region $0 < x < L_x$ and $0 < y < L_y$, the eigenmodes are of the form
\[ \Phi_{mn}(x,y) = \frac{1}{\mathscr{N}_{mn}}\cos(p_m x)\cos(q_n y), \]
where $\mathscr{N}_{mn} > 0$ is a normalisation constant.
Notably, the wavenumbers $p_m = m\pi/L_x$ and $q_n = n\pi/L_y$ are chosen so that the no-flux condition is satisfied (see equation \eqref{eq:Euler_no_flux_walls}).
For a triad determined by the non-negative integers $m_j$ and $n_j$ (for $j = 1, 2, 3$), the corresponding wavenumbers, $P_j = p_{m_j}$ and $Q_j = q_{n_j}$, must satisfy $P_1 + P_2 = P_3$ and $Q_1 + Q_2 = Q_3$ (under suitable reordering of the subscripts) in order for the eigenmode correlation condition \eqref{eq:corr_cond} to be satisfied. By defining the wave vector $\bm{k}_j = (P_j, Q_j)$, the conditions on $P_j$ and $Q_j$ simplify to the single requirement $\bm{k}_1 + \bm{k}_2 = \bm{k}_3$, where the triangle inequality supplies that $|\bm{k}_3| \leq |\bm{k}_1| + |\bm{k}_2|$. As the eigenvalues, $K_j^2$, of the negative Laplacian operator are related to the wave vectors via $K_j = |\bm{k}_j|$, we deduce that $K_3 \leq K_1 + K_2$. Owing to the violation of the left-hand bound in equation \eqref{eq:kineq}, we conclude that resonant triads \emph{cannot} exist in a rectangular cylinder of finite depth.

\subsubsection{Circular cylinder}
\label{sec:circular_cylinder}

We consider a circular cylinder of unit radius in dimensionless variables (i.e.\  the dimensional radius is equal to $a$; see \S \ref{sec:formulation}). For polar coordinates $\x = (r,\theta)$, it is well known that the corresponding (complex-valued) eigenmodes may be expressed in the form
\begin{equation}
\label{eq:Bessel_eig}
\Phi_{mn}(r,\theta) = \frac{1}{\mathscr{N}_{mn}}\Jb_m(k_{mn}r)\me^{\mi m \theta}, 
\quad\mathrm{where}\quad
\mathscr{N}_{mn} = \big|\Jb_m(k_{mn})\big|\sqrt{1 - \frac{m^2}{k_{mn}^2}} 
\end{equation}
is the normalisation factor and $m$ is the azimuthal wavenumber (an integer). Furthermore, the no-flux condition \eqref{eq:Euler_no_flux_walls} determines that the radial wavenumbers, denoted $k_{mn}$, satisfy $\Jb_m'(k_{mn}) = 0$, where $0 < k_{m1} < k_{m2} < \ldots$ (we exclude $k_{00} = 0$ from consideration; see \S \ref{sec:triads_existence}). Notably, the eigenvalues of the negative Laplacian operator are \emph{precisely} the squared wavenumbers, $k_{mn}^2$; consequently, the antinodes of each Bessel function play a pivotal role in determining the existence of resonant triads. 

Akin to the rectangular cylinder, we find that the eigenmode correlation condition imparts an important restriction on the combination of eigenmodes that may resonate. For given $m_j$ and $n_j$ (for $j = 1,2,3$), we denote $K_j = k_{m_jn_j}$, $\Psi_j = \Phi_{m_j n_j}$ and $N_j = \mathscr{N}_{m_j n_j}$. Although the correlation condition given in equation \eqref{eq:corr_cond} is defined for real eigenmodes, a similar condition holds for complex-valued eigenmodes, namely $\iint_{\mathcal{D}} \Psi_1 \Psi_2 \Psi_3^*\dA \neq 0$.
By considering the quantity
\[\iint_{\mathcal{D}} \Psi_1 \Psi_2  \Psi_3^* \dA =  \frac{1}{N_1 N_2 N_3}\bigg(\int_0^1 r\Jb_{m_1}(K_1 r) \Jb_{m_2}(K_2 r) \Jb_{m_3}(K_3 r)\dr\bigg)\bigg(\int_0^{2\pi} \me^{\mi(m_1 + m_2 - m_3)\theta}\dth\bigg),\] 
we deduce from the azimuthal integral that a necessary condition for the correlation integral to be nonzero is $m_1 + m_2 = m_3$ \cite{Michel2019}. This condition thus restricts the permissible combinations of azimuthal wavenumbers in a manner similar to the restriction on the permissible planar wavenumbers for the case of a rectangular cylinder. Unlike rectangular cylinders, however, we demonstrate that resonant triads \emph{are} possible in a circular cylinder.

\begin{table}
\begin{center}
\begin{tabular}{ c | ccc | ccc | cccc | c | c}
No.\ & $m_1$ & $m_2$ & $m_3$ & $n_1$ & $n_2$ & $n_3$ & $K_1$ & $K_2$ & $K_3$ & $\overline{K}$ &  $h_c$ & $\iint_{\mathcal{D}} \Psi_1\Psi_2\Psi_3^*\dA$ \\ 
\rowcolor{LightGrey}
1 & -1 & 1 & 0 & 1 & 1 & 2 & 1.841 & 1.841 & 7.016 & 3.566 & 1.00970 & -0.02032 \\
\rowcolor{Grey}
2 & 1 & 1 & 2 & 1 & 1 & 2 & 1.841 & 1.841 & 6.706 &  3.463 & 0.83138 & 0.02801 \\
3 & -1 & 2 & 1 & 1 & 1 & 3 & 1.841 & 3.054 & 8.536 & 4.477 & 0.60375 & -0.02595 \\
4 & 1 & 2 & 3 & 1 & 1 & 2 & 1.841 & 3.054 & 8.015 & 4.304 & 0.50595 & 0.03712 \\
5 & -1 & 3 & 2 & 1 & 1 & 3 & 1.841 & 4.201 & 9.969 & 5.337 & 0.48152 & -0.02717 \\
\rowcolor{LightGrey}
6 & -2 & 2 & 0 & 1 & 1 & 3 & 3.054 & 3.054 & 10.173 & 5.427 & 0.39129 & -0.03050 \\
7 & 0   & 1 & 1 & 1 & 1 & 3 & 3.832 & 1.841 & 8.536 & 4.736 & 0.38516 & 0.00542 \\
\rowcolor{DarkGrey}
8 & -1 & 1 & 0 & 1 & 2 & 3 & 1.841 & 5.331 & 10.173 & 5.782 & 0.30197 & -0.00603 \\
9 & 1 & 1 & 2 & 1 & 2 & 3 & 1.841 & 5.331 & 9.969 & 5.714 & 0.28691 & 0.01818 \\
10 & 0 & 2 & 2 & 1 & 1 & 3 & 3.832 & 3.054 & 9.969 & 5.619 & 0.26387 & -0.00087\\
11 & 1 & 2 & 3 & 1 & 2 & 3 & 1.841 & 6.706 & 11.346 & 6.631 & 0.23678 & 0.02590 \\
12 & 0 & 3 & 3 & 1 & 1 & 3 & 3.832 & 4.201 & 11.346 & 6.460 & 0.21395 &  -0.00640 \\
13 & 1 & 2 & 3 & 2 & 1 & 3 & 5.331 & 3.054 & 11.346 & 6.577 & 0.19839 & 0.01522 \\
\rowcolor{Grey}
14 &0   & 0 & 0 & 1 & 1 & 3 & 3.832 & 3.832 & 10.173 & 5.946 & 0.19814 & 0.03327 \\
15 & -1 & 2 & 1 & 1 & 1 & 2 & 1.841 & 3.054 & 5.331 & 3.409 & 0.17266 & 0.85581 \\
\rowcolor{LightGrey}
16 & -2 & 2 & 0 & 1 & 1 & 2 & 3.054 & 3.054 & 7.016 & 4.375 & 0.17030 & 0.46429 \\
17 & -1 & 3 & 2 & 1 & 1 & 2 & 1.841 & 4.201 & 6.706 & 4.250 & 0.16313 & 0.64211 \\
18 & -2 & 3 & 1 & 1 & 1 & 3 & 3.054 & 4.201 & 8.536 & 5.264 & 0.15767 & 0.30704 \\
\rowcolor{LightGrey}
19 & -1 & 1 & 0 & 1 & 1 & 1 & 1.841 & 1.841 & 3.832 & 2.505 & 0.15227 & 1.28795 \\
\rowcolor{LightGrey}
20 & -3 & 3 & 0 & 1 & 1 & 3 & 4.201 & 4.201 & 10.173 & 6.192 &  0.14591 & 0.19061 \\
21 & -2 & 2 & 0 & 1 & 2 & 3 & 3.054 & 6.706 & 10.173 & 6.645 & 0.06331 & 0.68257 \\
22 & -1 & 3 & 2 & 2 & 1 & 3 & 5.331 & 4.201 & 9.969 & 6.501 & 0.06286 & 0.66930 \\
23 & -1 & 2 & 1 & 2 & 1 & 3 & 5.331 & 3.054 & 8.536 & 5.641 & 0.04664 & 0.99088 \\
24 & -1 & 3 & 2 & 1 & 2 & 3 & 1.841 & 8.015 & 9.969 & 6.609 & 0.03928 & 1.08903 \\
25 & 0 & 3 & 3 & 2 & 1 & 3 & 7.016 & 4.201 & 11.346 & 7.521 & 0.02782 & 1.00669
\end{tabular}
\caption{\label{Table_circle} Combinations of the azimuthal wavenumbers, $m_j$, and radial mode indices, $n_j$, that form a resonant triad ($m_1 + m_2 = m_3$ and $\Omega_1 + \Omega_2 = \Omega_3$) at critical depth, $h_c$, in a circular cylinder of unit radius. For each triad, the corresponding wavenumbers, $K_j = k_{m_j n_j}$, satisfy \eqref{eq:kineq}, and the correlation condition,  $\iint_{\mathcal{D}} \Psi_1\Psi_2\Psi_3^*\dA \neq 0$, is met. The list is restricted to resonant triads arising for $|m_j|, n_j \leq 3$, and we consider $m_1 \leq m_2$ and $m_3 \geq 0$ without loss of generality. We have omitted resonances that give rise to the same critical depth, but with the roles of modes 1 and 2 swapped. The triad numbers (left column) and shaded rows are referenced in the text.
}
\end{center}
\end{table}

Despite the apparent restriction of the Bessel antinodes, $K_j$, and summation condition on the azimuthal wavenumbers, $m_j$, Theorem \ref{thm:triads} determines that a vast array of resonant triads may be excited for judicious choices of the fluid depth. In table \ref{Table_circle}, we list a small number of resonant triads and each corresponding critical depth, $h_c$, subject to the restrictions $|m_j| \leq 3$ and $n_j \leq 3$; for larger values of $|m_j|$ and $n_j$, the corresponding wave field becomes increasingly oscillatory, to the extent that the effects of surface tension and dissipation might become appreciable. 
Moreover, even marginally relaxing the upper bounds on $|m_j|$ and $n_j$ vastly increases the number of resonant triads; indeed, the restriction $|m_j| \leq 4$ and $n_j \leq 4$ introduces 70 additional resonant triads relative to table \ref{Table_circle}. As the upper bounds for $|m_j|$ and $n_j$ are further increased, the typical difference between the various critical depths decreases \refone{and an increasingly large number of triads form at small values of the critical depth. Triads forming in shallow fluids (e.g.\ triads 21 to 25 in table \ref{Table_circle}) have physical relevance only at larger length scales (e.g.\ lakes) as dissipation could become a dominant factor at smaller scales.
}

\refone{Although the list of triads in table \ref{Table_circle} is restricted to the lowest radial and azimuthal modes, we observe some general trends. In particular, we observe that the correlation integral, $\iint_{\mathcal{D}} \Psi_1\Psi_2\Psi_3^*\dA$, generally decreases in magnitude as the fluid depth increases. Although the correlation integral remains non-zero (as is necessary to satisfy the correlation condition), its small value in some cases (e.g.\ triad 10) potentially corresponds to an elongation of the triad evolution time-scale (see \S \ref{sec:multiple_scales}). Moreover, we observe that the average wavenumber involved in the triad, $\overline{K} = \frac{1}{3}(K_1 + K_2 + K_3)$, is appreciably larger when the critical depth is very small. This correlation is consistent with the form of the corresponding angular frequency, $\Omega_j = \sqrt{K_j \tanh(K_j h_c)}$, for which a small critical depth, $h_c$, is necessary for finite-depth effects to be appreciable when the typical wavenumber is large. To enumerate the myriad resonant triads arising in a circular cylinder when the upper bounds on $|m_j|$ and $n_j$ are relaxed, we provide MATLAB code in the supplementary material.}

At this juncture, it is informative to assess how the triads listed in table \ref{Table_circle} relate to the resonances explored in prior investigations. First,  \refone{triad 8} in table \ref{Table_circle} (dark grey row) was explored by Michel \cite{Michel2019} for a circular cylinder of radius 9.45 cm and an approximate fluid depth of 3 cm; it follows that the depth-to-radius ratio in Michel's experiment was approximately 0.317, close to the value of 0.30197 reported in table \ref{Table_circle}.
Furthermore, table \ref{Table_circle} (grey rows) incorporates two well-known examples of a 1:2 resonance, for which modes 1 and 2 coincide: (i) the critical depth $h_c = 0.83138$ \refone{(triad 2)} corresponds to the second-harmonic resonance with the fundamental mode \cite{Miles1976, Miles1984a, Bryant1989, Yang2021}; (ii) the critical depth $h_c = 0.19814$ \refone{(triad 14)} corresponds to a standing wave composed of two resonant axisymmetric modes \cite{Mack1962, Yang2021}.
Finally, \refone{triads 1, 6, 16, 19 and 20}  (table \ref{Table_circle}, light grey rows) form an interesting class of resonant triad, for which an axisymmetric mode ($m_3 = 0$) interacts with two identical counter-propagating non-axisymmetric modes ($m_1= -m_2 \neq 0$ and $n_1 = n_2$). In fact, our investigation in \S \ref{sec:pump_modes} demonstrates that the axisymmetric mode is the so-called \emph{pump} mode, and may thus excite the non-axisymmetric modes, even when the initial energy in each non-axisymmetric mode is negligible. We draw an analogy between this novel class of resonant triad and the excitation of beach edge waves \cite{Guza1974} in \S \ref{sec:discussion}.

We conclude our exploration of resonant triads arising in a circular cylinder by remarking that the fluid depth may, in some cases, be judiciously chosen so as to excite multiple triads. In general, the condition on the angular frequencies, $\Omega_1 + \Omega_2 = \Omega_3$ (see equation \eqref{eq:triad_sum1}), cannot be satisfied for two distinct triads at the same fluid depth; however, nonlinear resonance may persist for both triads provided that each condition on the angular frequencies is \emph{approximately} satisfied \cite{Bretherton1964, McGoldrick1972, CraikBook}, at the cost of weak detuning (see \S \ref{sec:weak_detuning} for further details). Specifically, if triads 1 and 2 have critical depths $h_{c,1}$ and $h_{c,2}$, respectively, then there is potential excitement of both triads when the fluid depth, $h$, satisfies $|h - h_{c,j}| = O(\epsilon)$ for $j = 1,2$ (where $0 < \epsilon \ll 1$ is the typical wave slope; see \S \ref{sec:formulation}), giving rise to the approximation $\Omega_1 + \Omega_2 - \Omega_3 = O(\epsilon)$ for each triad. For example, if $0 < h_{c,2} - h_{c,1} \ll 1$, then it may be sufficient to excite both triads at an intermediate depth, $h_{c,1} \leq h \leq h_{c,2}$. We note, however, that the excitation of multiple triads at a single fluid depth is not possible when the depth discrepancy, $|h - h_{c,j}|$, becomes too large (relative to the typical wave slope) for any of the triads under consideration.

To demonstrate the potential for the simultaneous excitation of two triads within a circular cylinder of finite depth, we consider two scenarios: (i) the excitation of two triads that share a common wave mode; and (ii) the excitation of two triads that do not share any common wave modes. Heuristically, case (ii) is more common than case (i) owing to the number of similar fluid depths in table \ref{Table_circle}; however, case (i) will likely generate a far richer set of dynamics owing to the nonlinear interaction between the two triads \cite{McEwan1972, CraikBook, Chow1996, Choi2021}. As an example of case (i), we consider \refone{triads 11 and 12} in table \ref{Table_circle}, with nearby critical depths \refone{$h_{c,1} = 0.23678$ and $h_{c,2} = 0.21395$}, respectively. As mode $(m_3,n_3) = (3,3)$ is common to both triads, inter-triad resonance may arise at an intermediate depth, e.g.\ $h = 0.225$.
Furthermore, an example of case (ii) arises for \refone{triads 13 and 14} in table \ref{Table_circle}, with nearby critical depths \refone{$h_{c,1} = 0.19839$ and $h_{c,2} = 0.19814$}. Neither of these triads share a common wave mode, so one would \emph{not} expect the inter-triad energy exchange discussed in case (i). Nevertheless, one might anticipate a signature of these two triads to be visible in the  surface evolution for an intermediate depth, e.g.\ $h = 0.19825$. The theoretical and numerical exploration of coupled triads in a circular cylinder will be the focus of future investigation.

\subsubsection{Annular cylinder}
\label{sec:annular_cylinder}

A natural variation upon a circular cylinder is an annulus of inner radius $r_0 \in (0,1)$ and outer radius 1. By varying $r_0$, the annulus approaches a circular cylinder as $r_0 \rightarrow 0^+$, and a quasi-one-dimensional periodic ring as $r_0 \rightarrow 1^-$. Notably, resonant triads are impossible for a one-dimensional periodic ring, as can be shown by modifying the arguments presented for the case of a rectangular cylinder (see \S \ref{sec:rectangular_cylinder}). Thus, one might anticipate that the existence of triads in an annular cylinder depends critically on the inner radius, $r_0$. Rather than enumerating some possible triads for given values of $r_0$, we instead track the corresponding critical depth, $h_c$, for the triads identified for a circular cylinder (see table \ref{Table_circle}) as $r_0$ is progressively increased from zero. Of particular interest is determining whether a given triad exists for all $r_0 < 1$, or whether there is some critical inner radius, $r_c$, beyond which the triad ceases to exist, with either $h_c \rightarrow 0$ or $h_c \rightarrow \infty$ as $r_0 \rightarrow r_c^-$.

The (complex-valued) eigenmodes in an annular domain are cylinder functions of the form
\begin{equation}
\label{eq:cylinder_functions}
\Phi_{mn}(r,\theta) = \frac{1}{\mathscr{N}_{mn}}\bigg[\Jb_m(k_{mn}r) \cos(\gamma_{mn}\pi) + \Yb_m(k_{mn}r) \sin(\gamma_{mn}\pi) \bigg]  \me^{\mi m \theta},
\end{equation}
where $\mathscr{N}_{mn} > 0$ is a normalisation constant, $\Yb_m$ is the Bessel function of the second kind with order $m$ (an integer), and $\gamma_{mn} \in [0,1]$ determines the weighting between the two Bessel functions. As shown in appendix \ref{app:annulus_wavenumbers}, the no-flux condition (see equation \eqref{eq:Euler_no_flux_walls}) on the inner and outer walls determines that the wavenumbers, $k_{mn}(r_0)$, satisfy the equation
\begin{equation}
\label{eq:annulus_no_flux}
\Jb_m'(k_{mn} r_0) \Yb_m'(k_{mn}) - \Jb_m'(k_{mn}) \Yb_m'(k_{mn} r_0)  = 0.
\end{equation}
A formula for the corresponding value of $\gamma_{mn}$ is determined in appendix \ref{app:annulus_wavenumbers}.  Once again, the wavenumbers, $k_{mn}$, are ordered so that $0 < k_{m1} < k_{m2} < \ldots$ (excluding $k_{00} = 0$) and satisfy $-\Delta \Phi_{mn} = k_{mn}^2\Phi_{mn}$. Three correlated wave modes may form a resonant triad (for a judicious choice of the fluid depth) provided that the corresponding wavenumbers, $K_j$, which depend on the channel width, $1 - r_0$, satisfy the bounds given in Theorem \ref{thm:triads}.

\begin{figure}
\centering
\includegraphics[width=1\textwidth]{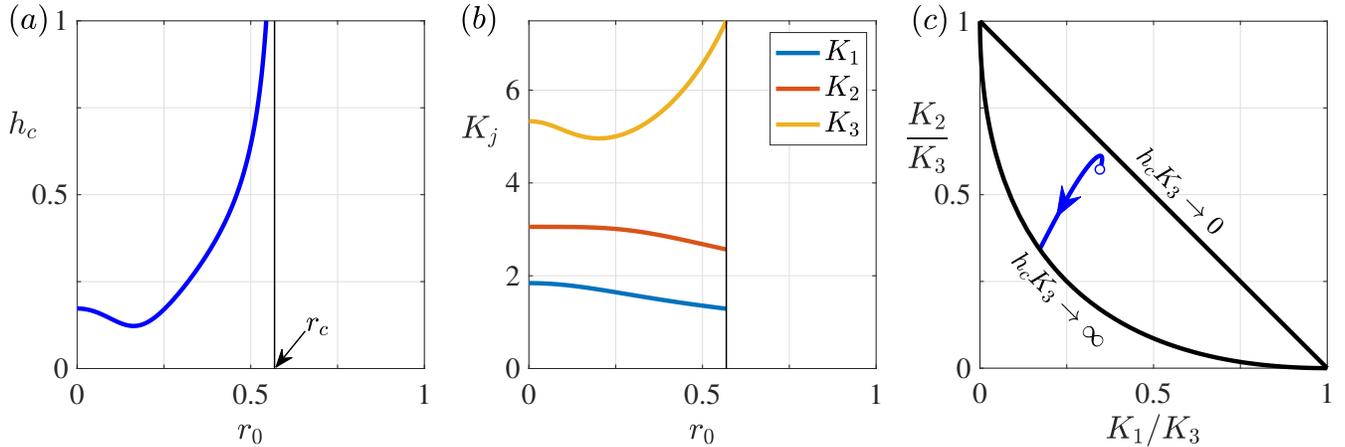}
\caption{\label{fig:Annulus_example}
The existence and predominant characteristics of a triad in an annular cylinder with inner radius $r_0$ and outer radius 1. The triad bifurcates from the critical depth $h_c = 0.17266$ as $r_0 \rightarrow 0$ (the limiting case of a circular cylinder), with corresponding wavenumbers presented in table \ref{Table_circle} (see \refone{triad 15}). 
$(a)$ The critical depth, $h_c$ (blue curve), with $h_c \rightarrow \infty$ as $r_0 \rightarrow r_c^-$, where $r_c \approx 0.57$ (black line). 
$(b)$ The corresponding wavenumbers, $K_j$, all of which remain finite for $r_0 < r_c$ (black line).
$(c)$ The normalised wavenumbers, $K_1/K_3$ and $K_2/K_3$, parametrised by increasing $r_0$ (blue arrow), with the limiting case $r_0 \rightarrow 0$ denoted by the white dot. The wavenumbers leave the triad existence region (see Theorem \ref{thm:triads}) via the left-hand boundary (black curve) as $r_0 \rightarrow r_c^-$.
}
\end{figure}
Bifurcating from the limiting case of a circular cylinder, we track the critical depth (when such a depth exists) of different triads as $r_0$ is progressively increased. The predominant behaviour is characterised by the example presented in figure \ref{fig:Annulus_example}, for which we consider the triad whose critical depth is $h_c = 0.17266$ as $r_0 \rightarrow 0^+$ (see \refone{triad 15} in table \ref{Table_circle}). Given that $h_c$ is fairly small in this limit, one might anticipate that the triad ceases to exist with $h_c \rightarrow 0$; somewhat surprisingly, however, the opposite scenario arises, with $h_c \rightarrow \infty$ as $r_0 \rightarrow r_c^- $ ($r_c \approx 0.57$ in this example). It follows, therefore, that the triad may persist for narrow channels only when the fluid is sufficiently deep. We note, however, that there exist (at least) two relatively rare transitions for increasing $r_0$, which we briefly describe as follows: (i) the triad ceases to exist when $h_c \rightarrow 0$ as $r_0 \rightarrow r_c^-$, which may arise when bifurcating from a sufficiently shallow circular cylinder (e.g.\ \refone{triad 25} in table \ref{Table_circle}); and (ii) the triad continues to exist for all $r_0 < 1$, with $h_c \rightarrow 0$ and $K_j \rightarrow \infty$ as $r_0 \rightarrow 1$, yet the normalised depth, $h_c K_3$, remains finite, and the normalised wavenumbers, $K_1/K_3$ and $K_2/K_3$, remain within the triad existence region (e.g.\ \refone{triad 14} in table \ref{Table_circle}). Owing to the appreciable influence of viscous effects for relatively shallow fluids, the physical relevance of these latter two scenarios is somewhat nebulous, however.

\section{The evolution of resonant triads}
\label{sec:triad_eqs}

Having established the existence of resonant triads, we now determine the long-time triad evolution, utilising the method of multiple scales. Ostensibly, the calculations necessary for determining the triad equations are a variation upon the pioneering work of McGoldrick \cite{McGoldrick1965, McGoldrick1970b, McGoldrick1970a} in the absence of surface tension. However, the confinement of the fluid to a cylinder imposes some additional considerations, the salient details of which we outline below. Finally, we note that an alternative approach to multiple scales is Whitham's technique of averaging the system's Lagrangian \cite{Whitham1965b, Whitham1965a, Whitham1967a, Whitham1967b}, which has the advantage of streamlining some algebraic calculations \cite{Simmons1969, Miles1976, Miles1984a}; nevertheless, multiple-scales analysis is sufficient for our purposes and allows for the possible inclusion of higher-order corrections in the asymptotic expansion \cite{McGoldrick1970a}.

In a manner similar to \S \ref{sec:triads_existence}, we consider three linear wave modes (with real-valued eigenfunctions), enumerated $n_1$, $n_2$ and $n_3$, where we denote
\[\Omega_j = \omega_{n_j}, \quad K_j = k_{n_j}, \quad L_j = \DtNhat_{n_j}, \quad \mathrm{and}\quad \Psi_j(\x) = \Phi_{n_j}(\x) \quad \mathrm{for}\,\,\, j = 1,2,3. \]
In contrast to \S \ref{sec:triads_existence}, however, we now allow each (nonzero) angular frequency to be either negative or positive: the resonance condition on the angular frequencies is henceforth defined
\begin{equation}
\label{eq:omega_sum2}
\Omega_1 + \Omega_2 + \Omega_3 = 0.
\end{equation}
The modified requirement on the angular frequencies (equation \eqref{eq:omega_sum2}) is not restrictive on the possible triad combinations; one may recover equation \eqref{eq:triad_sum1} by mapping $\Omega_3 \mapsto -\Omega_3$, for example. The decision behind the summation condition on the angular frequencies is motivated by the cyclical symmetry of equation \eqref{eq:omega_sum2}, a property that will be inherited by the resultant amplitude equations \cite{Simmons1969}. As a consequence, one need only derive the amplitude equation for one of the wave modes; the amplitude equations for the remaining two wave modes follow by cyclic permutation of the subscripts $(1,2,3)$. 

Before embarking on the multiple-scales analysis presented in \S \ref{sec:multiple_scales}, we remark upon two caveats.
First, we note that equation \eqref{eq:omega_sum2} corresponds to an exact resonance, for which the fluid depth, $h$, is chosen to be precisely equal to the critical depth, $h_c$. In practice, however, there may be a small discrepancy between $h$ and $h_c$, resulting in a the sum of the angular frequencies being slightly offset from zero. When the frequency detuning is sufficiently weak, e.g.\ $\Omega_1 + \Omega_2 + \Omega_3 = O(\epsilon)$, one may modify the following asymptotic analysis to derive a similar set of amplitude equations (see \S \ref{sec:weak_detuning}).
Second, our analysis in \S \ref{sec:multiple_scales} is not valid when two of the wave modes coincide. This case corresponds to a 1:2 resonance, for which the corresponding evolution equations were derived by Miles \cite{Miles1976} using Whitham modulation theory (as summarised in \S \ref{sec:12_resonance}).

\subsection{Multiple-scales analysis}
\label{sec:multiple_scales}

In order to determine the evolution of each of the three dominant wave modes involved in an exact resonance, we utilise the method of multiple scales \cite{KevorkianBook, Strogatz}. Specifically, we seek a perturbation solution to the Benney-Luke equation \eqref{eq:BL_eq} of the form $u \sim u_0 + \epsilon u_1 + O(\epsilon^2)$. The leading-order terms in equation \eqref{eq:BL_eq} determine that $u_0$ satisfies $\partial_{tt}u_0 + \DtN u_0 = 0$; we \emph{choose} to consider a leading-order solution comprised only of the three triad modes (all other modes are assumed to be smaller in magnitude and appear at higher order), giving rise to the leading-order form
\begin{equation}
\label{eq:triad_leading_order}
u_0(\x ,t, \tau) = \sum_{j = 1}^3 \Big[ A_j(\tau) \Psi_j(\x) \me^{-\mi \Omega_j t} + \cc\Big].
\end{equation}
In equation \eqref{eq:triad_leading_order}, we have introduced the slow time-scale $\tau = \epsilon t$, which governs the evolution of each complex amplitude, $A_j$. As $\epsilon$ and $t$ are both independent variables, we treat $\tau$ and $t$ as independent time-scales, giving rise to the transformation of derivatives $\partial_t \mapsto \partial_t + \epsilon \partial_\tau$. Finally, $\cc$ denotes the complex conjugate of the preceding term, a contribution necessary for real $u_0$.

So as to determine coupled evolution equations for each complex amplitude, $A_j$, we consider terms of $O(\epsilon)$ in the Benney-Luke equation \eqref{eq:BL_eq}. By substituting the leading-order solution, $u_0$, into the nonlinear terms and applying the triad condition for the angular frequencies (equation \eqref{eq:omega_sum2}), we obtain the following problem for $u_1$:
\begin{equation}
\label{eq:triad_u1}
\partial_{tt}u_1 + \DtN u_1 = -\bigg[\sum_{j = 1}^3 f_j(\x,\tau)\me^{-\mi \Omega_j t} + \cc\bigg] + \mbox{nonresonant terms}.
\end{equation}
As we will see below, each of the functions $f_j(\x,\tau)$ appearing on the right-hand side of equation \eqref{eq:triad_u1} will play a fundamental role when determining the amplitude equations; specifically,
\[
f_1 = -2\mi \Omega_1 \sd{A_1}{\tau} \Psi_1 + \mi A_2^* A_3^*\bigg( \Big[\Omega_2 \big(L_3^2 - K_3^2\big) + \Omega_3\big(L_2^2 - K_2^2\big) - 2\Omega_1 L_2L_3\Big] \Psi_2\Psi_3 - 2\Omega_1 \nabla\Psi_2 \cdot \nabla \Psi_3\bigg),
\]
where $f_2$ and $f_3$ follow upon cyclic permutation of the subscripts $(1,2,3)$. Finally, we note that the `nonresonant terms' in equation \eqref{eq:triad_u1} are of the general form $p(\x,\tau)\me^{\mi \varsigma t}$, where we assume that the angular frequency, $\varsigma$, is not equal (or close) to any of the angular frequencies, $\pm\omega_n$, associated with linear wave modes (see \S \ref{sec:triads_existence}). 

We proceed by projecting equation \eqref{eq:triad_u1} onto each of the three wave modes, giving rise to differential equations of the form (for $j = 1,2,3$)
\begin{equation}
\label{eq:uhat}
\partial_{tt}\hat{u}_{1,j} + L_j \hat{u}_{1,j} = -\Big[\langle \Psi_j, f_j \rangle \me^{-\mi \Omega_j t} + \cc \Big] + \mbox{nonresonant terms},
\end{equation}
where $\hat{u}_{1,j} = \langle \Psi_j, u_1 \rangle $ is the projection of $u_1$ onto the mode $\Psi_j$. By recalling that $L_j = \Omega_j^2$, we immediately see that the term in square brackets in equation \eqref{eq:uhat} is itself a solution to the linear operator $\partial_{tt} + \Omega_j^2$. It follows that the solution of equation \eqref{eq:uhat} comprises of particular solutions that have temporal dependence $t\me^{\pm\mi \Omega_j t}$, leading to an ill-posed asymptotic expansion when $\epsilon t = O(1)$. The resolution to this problem is achieved via the solubility condition 
$\langle \Psi_j, f_j\rangle = 0, $
which suppresses the secular growth.

By applying the solubility condition $\langle \Psi_j, f_j\rangle = 0$ for $j = 1,2,3$, we conclude that the complex amplitude, $A_j(\tau)$, of each wave mode, $\Psi_j(\x) \me^{-\mi\Omega_j t}$, evolves according to the triad system of canonical form \cite{Bretherton1964, CraikBook}
\begin{equation}
\label{eq:triad_equations}
\sd{A_1}{\tau} = \alpha_1 A_2^* A_3^*, \qquad
\sd{A_2}{\tau} = \alpha_2 A_1^* A_3^*, \qquad
\sd{A_3}{\tau} = \alpha_3 A_1^* A_2^*,
\end{equation}
where
\begin{equation}
\label{eq:alpha1_full}
\alpha_1 = \frac{1}{2\Omega_1}\bigg( \Big[\Omega_2 \big(L_3^2 - K_3^2\big) + \Omega_3\big(L_2^2 - K_2^2\big) - 2\Omega_1 L_2L_3\Big] \mathscr{C} -  2\Omega_1 \big\langle \Psi_1,  \nabla\Psi_2 \cdot \nabla \Psi_3 \big\rangle\bigg),
\end{equation}
while $\alpha_2$ and $\alpha_3$ follow by cyclic coefficient of the subscripts $(1,2,3)$.
Furthermore, the correlation integral, $\mathscr{C}$, is defined
\begin{equation}
\label{eq:corr_int_def}
\mathscr{C} = \frac{1}{S}\iint_\mathcal{D} \Psi_1\Psi_2\Psi_3\dA,
\end{equation}
where we recall that $S$ is the area of the cylinder cross-section (see \S \ref{sec:DtN_operator}).
As the triad equations \eqref{eq:triad_equations} are valid for $\tau = O(1)$ (or $t = O(1/\epsilon)$), their dynamics yield an informative view of the long-time evolution of the resonant triad.

In order to assess the influence of the triad coefficients on the triad evolution (see \S \ref{sec:simplify_coeff}), we first simplify the algebraic form given in equation \eqref{eq:alpha1_full}.
As shown by Miles \cite{Miles1976}, one may simplify the inner product $\langle \Psi_1, \nabla \Psi_2 \cdot \nabla \Psi_3 \rangle$ by repeated application of the divergence theorem and utilisation of the relationship $-\Delta \Psi_j = K_j^2\Psi_j$; it follows that
\begin{equation}
\label{eq:Miles_simp}
\big\langle \Psi_1, \nabla\Psi_2 \cdot \nabla \Psi_3 \big\rangle = \frac{1}{2}\Big(K_2^2 + K_3^2 - K_1^2\Big) \mathscr{C},
\end{equation}
where $\mathscr{C}$ is the correlation integral defined in equation \eqref{eq:corr_int_def}.
We then substitute equation \eqref{eq:Miles_simp} into equation \eqref{eq:alpha1_full} and simplify using the relation $\Omega_1 + \Omega_2 + \Omega_3 = 0$. After some algebra, we derive the reduced expression
\begin{equation}
\label{eq:alpha1_simp}
\alpha_1 = \frac{\mathscr{C}}{2\Omega_1}\bigg( \Omega_2 L_3^2 + \Omega_3L_2^2 - 2\Omega_1 L_2 L_3 + \sum_{l = 1}^3 \Omega_l K_l^2  \bigg),
\end{equation}
where $\alpha_2$ and $\alpha_3$ follow similarly.
Finally, we demonstrate in appendix \ref{app:reduce_triad_coeff} that the algebraic form of the triad coefficients may be further reduced to
\begin{equation}
\label{eq:triad_coeff_final}
\alpha_j = \frac{ \mathscr{C}\beta}{2\Omega_j} \quad\mathrm{for}\quad j = 1,2,3,
\end{equation}
where
\begin{equation}
\label{eq:beta_coeff}
\beta = \sum_{l= 1}^3 \Omega_lK_l^2 - \frac{1}{2}\Omega_1\Omega_2\Omega_3 \big(\Omega_1^2 + \Omega_2^2 + \Omega_3^2\big).
\end{equation}
Equations \eqref{eq:triad_equations}, \eqref{eq:corr_int_def}, \eqref{eq:triad_coeff_final} and \eqref{eq:beta_coeff} constitute the triad equations for resonant gravity waves confined to a cylinder of finite depth. Although the triad equations are of canonical form \cite{Bretherton1964}, the novelty of our investigation is the computation of the coefficients, $\alpha_j$, whose algebraic form is specific to our system.

\subsection{Properties of the triad coefficients}
\label{sec:simplify_coeff}

The simplified form of the coefficients, $\alpha_j$ (equation \eqref{eq:triad_coeff_final}), allows for some important theoretical observations that were obfuscated by the more complicated expressions for $\alpha_j$ given in equations \eqref{eq:alpha1_full} and \eqref{eq:alpha1_simp}.
In particular, as exactly two of the angular frequencies, $\Omega_j$, have the same sign, we deduce from equation \eqref{eq:triad_coeff_final} that the two corresponding coefficients, $\alpha_j$, also have the same sign, with the third coefficient having the opposite sign. By utilising well-known results pertaining to the canonical triad equations, we conclude that all solutions to the triad equations \eqref{eq:triad_equations} are periodic in time, with solutions expressible in terms of elliptic functions \cite{Ball1964, Bretherton1964, Simmons1969, CraikBook}. Typically, these solutions result in an exchange of energy between the comprising modes, although there is a class of periodic solution that, perhaps counter-intuitively, results in zero energy exchange for all time \cite{CaseChiu1977, ChabaneChoi2019}.
Moreover, it is readily verified that the leading-order energy density, $\mathscr{E}_1 + \mathscr{E}_2 + \mathscr{E}_3$, is conserved, where $\mathscr{E}_j = \Omega_j^2 |A_j|^2$, consistent with the Hamiltonian structure of the Euler equations \cite{Bretherton1964, CraikBook}. 
The reader is directed to the work of Craik \cite{CraikBook} for a more detailed account of the various properties of the canonical triad equations.
\begin{figure}
\centering
\includegraphics[width=0.65\textwidth]{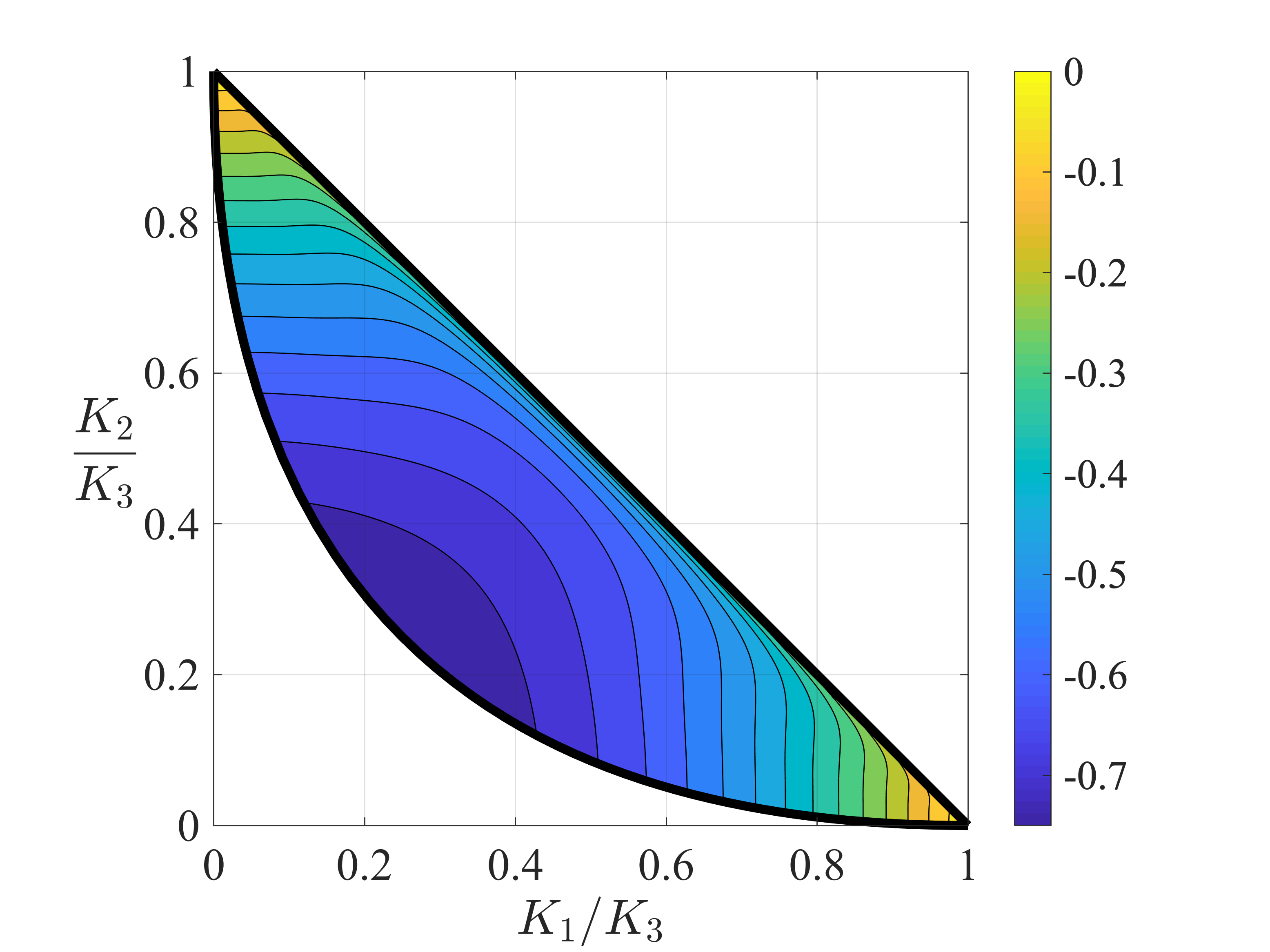}
\caption{\label{fig:beta_coefficient} Contours of $\beta/K_3^2$ (see equation \eqref{eq:beta_coeff}) for the case $\Omega_1, \Omega_2 > 0$ and $\Omega_3 < 0$ (with $\Omega_1 + \Omega_2 + \Omega_3 = 0$), for which $\beta < 0$ (see equation \eqref{eq:beta_bound}).
}
\end{figure}

Of particular relevance to the evolution of the triad is the quantity $\beta$ (see equation \eqref{eq:beta_coeff}), which, together with $\mathscr{C}$, determines the time scale over which energy exchange arises.
In particular, we present the form of $\beta$ in figure \ref{fig:beta_coefficient} for the case $\Omega_1,\Omega_2 > 0$ and $\Omega_3 < 0$. As we will demonstrate below, $\beta < 0$ in this case; 
in general, the sign of $\beta$ is the same as the sign of the largest (in magnitude) angular frequency, $\Omega_j$. 
Notably, $|\beta|$ decreases sharply towards zero as $K_1 + K_2 \rightarrow K_3$, corresponding to the limit $h_c \rightarrow 0$.
Similarly, $|\beta|$ approaches zero in the limiting cases $K_1 \ll K_3$ or $K_2 \ll K_3$, corresponding to one low-oscillatory wave mode interacting with two highly-oscillatory wave modes. 
Away from these limiting cases, however, $|\beta|$ depends only weakly on the wavenumbers, $K_j$, suggesting that the correlation integral, $\mathscr{C}$, predominantly controls the time-scale of the triad evolution.
Finally, we observe that $\beta$ is symmetric about the line $K_1 = K_2$, consistent with the invariance of equation \eqref{eq:beta_coeff} under the mapping $K_1\leftrightarrow K_2$ (and hence, $\Omega_1 \leftrightarrow \Omega_2$).

We conclude this section by proving that $\beta < 0$ in the case $\Omega_1, \Omega_2 > 0$ and $\Omega_3 < 0$. 
By comparing the forms of equations \eqref{eq:beta_coeff} and \eqref{eq:alpha1_simp}, and then permuting the subscripts $(1,2,3) \mapsto (3,1,2)$, we first note that $\beta$ may be equivalently expressed as
\[
\beta = \Omega_1 L_2^2 + \Omega_2L_1^2 - 2\Omega_3L_1L_2 + \sum_{l = 1}^3\Omega_l K_l^2, 
\]
or 
\[
\beta = \Omega_1 (L_2^2 + K_1^2) + \Omega_2(L_1^2 + K_2^2) + |\Omega_3| ( 2L_1 L_2 - K_3^2).
\]
By bounding $L_j = K_j \tanh(K_j h_c) < K_j$ for $0 < h_c < \infty$ and utilising the relation $\Omega_1 + \Omega_2 = |\Omega_3|$, we obtain 
\begin{equation}
\label{eq:beta_bound}
\beta < |\Omega_3|\big(K_1^2 + K_2^2 + 2K_1 K_2 - K_3^2\big) = |\Omega_3|\big((K_1 + K_2)^2 - K_3^2\big). 
\end{equation}
As resonant triads exist only when $K_1 + K_2 < K_3$ (see Theorem \ref{thm:triads}), we conclude that $\beta < 0$ in this case.

\subsection{Summary}
\label{sec:multiple_scales_summary}

To summarise our theoretical developments, the velocity potential, $u$, at the fluid rest level ($z = 0$) evolves according to
\begin{equation}
\label{eq:u_triads_exp}
u(\x,t) \sim \sum_{j = 1}^3 \Big[ A_j(\tau) \Psi_j(\x) \me^{-\mi \Omega_j t} + \cc\Big] + O(\epsilon), 
\end{equation}
where the complex amplitudes, $A_j(\tau)$, evolve over the slow time-scale, $\tau = \epsilon t$, according to the triad equations \eqref{eq:triad_equations}. In particular, the triad coefficients, $\alpha_j$ (see equation \eqref{eq:triad_coeff_final}), are defined in terms of the correlation integral, $\mathscr{C}$ (equation \eqref{eq:corr_int_def}), and the coefficient $\beta$ (equation \eqref{eq:beta_coeff}). Notably, we assume that $\mathscr{C}$ is nonzero; if this condition were violated then all three of the triad coefficients, $\alpha_j$, would be equal to zero, giving rise to non-interacting wave modes at leading order (contradicting the notion of a triad). Indeed, the condition $\mathscr{C} \neq 0$ is identical to the correlation condition detailed in equation \eqref{eq:corr_cond}, the origins of which we have now justified. Finally, the evolution of the free surface, $\eta$, may be recovered by recalling that $\eta = -u_t + O(\epsilon)$: we conclude that $\eta(\x,t)$ has a similar leading-order form to $u(\x,t)$, but each complex amplitude, $A_j(\tau)$, in \eqref{eq:u_triads_exp} is replaced by $\mi \Omega_j A_j(\tau)$ (see equation \eqref{eq:eta_exp} below).

We briefly contrast our investigation of triad interaction with the early-time calculation of Michel \cite{Michel2019}, who characterised the initial linear growth of a child mode induced by the nonlinear interaction of two parent modes (where all three modes comprise the triad). 
If modes 1 and 2 are the parent modes and mode 3 is the child mode, then the initial linear growth may be deduced directly from triad equations \eqref{eq:triad_equations} in the limit $|A_3| \ll |A_1| \sim |A_2|$. Specifically, the initial variation of $A_1$ and $A_2$ is slow relative to that of $A_3$, which has the approximate early-time form $A_3(\tau) \approx \alpha_3 C_1^* C_2^* \tau + C_3$, where $C_j = A_j(0)$. Notably, the linear growth rate of the child mode depends on the corresponding triad coefficient, $\alpha_3$, and the product of the initial amplitudes of the two parent modes.
However, our result for circular cylinders differs to that of Michel; we believe that the author neglected some important nonlinear contributions (compare Michel's equation (A2) to equations (2.4) and (2.4a) of Longuet-Higgins \cite{LonguetHiggins1962}). As Michel's experiment verified the scaling of the interaction only up to a proportionality constant,  this discrepancy was not captured.

\subsubsection{The influence of weak detuning}
\label{sec:weak_detuning}

As discussed earlier in \S \ref{sec:triad_eqs}, the analysis in \S\S \ref{sec:multiple_scales} and \ref{sec:simplify_coeff} does not account for weak detuning of the angular frequencies, as might arise when the fluid depth, $h$, differs slightly from the critical depth, $h_c$. We now briefly consider the case of weak detuning, for which equation \eqref{eq:omega_sum2} is replaced by the condition $\Omega_1 + \Omega_2 + \Omega_3 = \epsilon \sigma$ (see \S \ref{sec:circular_cylinder}); here $\epsilon$ is the small parameter representative of the typical wave slope (see \S \ref{sec:formulation}) and $\sigma = O(1)$ determines the extent of the detuning \cite{Bretherton1964, McGoldrick1972}. By following a very similar multiple-scales procedure to the case $\sigma = 0$, we obtain amplitude equations that are now augmented by a time-dependent modulation. Specifically, each complex amplitude now evolves according to
\[\sd{A_1}{\tau} = \alpha_1 A_2^* A_3^* \me^{\mi \sigma \tau}, \quad
\sd{A_2}{\tau} = \alpha_2 A_1^* A_3^* \me^{\mi \sigma \tau}, \quad
\sd{A_3}{\tau} = \alpha_3 A_1^* A_2^* \me^{\mi \sigma \tau}, \]
where each coefficient, $\alpha_j$, is defined in equation \eqref{eq:triad_coeff_final}. Although detuning yields non-autonomous amplitude equations, autonomous equations may be derived by mapping $A_j(\tau) \mapsto A_j(\tau) \me^{\mi \sigma\tau/3}$ for all $j = 1,2,3$ \cite{CraikBook}. Finally, we note that the energy, $\mathscr{E}_1 + \mathscr{E}_2 + \mathscr{E}_3$, is not exactly conserved when considering the effects of detuning; instead, the energy slowly oscillates about a constant value \cite{CraikBook}.

\subsubsection{The case of a 1:2 resonance}
\label{sec:12_resonance}

A 1:2 resonance is a resonant triad for which two modes comprising the triad coincide. For this case, we define two angular frequencies, $\Omega_1$ and $\Omega_2$, so that $\Omega_2 = 2\Omega_1$ \cite{Miles1976}, where the connection to resonant triads is clear when writing $\Omega_1 + \Omega_1 = \Omega_2$. By following a very similar multiple-scales procedure to that outlined in \S \ref{sec:multiple_scales}, we obtain
\[ u(\x,t) \sim \sum_{j = 1}^2 \Big[A_j(\tau) \Psi_j(\x) \me^{-\mi \Omega_j t} + \cc\Big] + O(\epsilon), \]
where 
\begin{equation}
\label{eq:Wilton_ripples_amplitude_eqs}
\sd{A_1}{\tau} = -\gamma A_1^* A_2 
\quad \mathrm{and}\quad 
\sd{A_2}{\tau} = \frac{\gamma}{4} A_1^2.
\end{equation}
In particular, the evolution of the amplitude equations \eqref{eq:Wilton_ripples_amplitude_eqs} depends on the coefficient $\gamma = \mathscr{C}\big(K_2^2 - K_1^2 - 3\Omega_1^4\big)$, where $\mathscr{C} = \frac{1}{S}\iint_{\mathcal{D}} \Psi_1^2 \Psi_2 \dA$ is the correlation integral. Indeed, the amplitude equations \eqref{eq:Wilton_ripples_amplitude_eqs} and coefficient, $\gamma$, are consistent with the results of Miles \cite{Miles1976} when expressing the evolution of each complex amplitude, $A_j$, in polar form (with appropriate rescaling). Finally, we note that a weak detuning (see \S \ref{sec:weak_detuning}) may also be incorporated within the amplitude equations \eqref{eq:Wilton_ripples_amplitude_eqs}, thereby accounting for a slight mismatch between the fluid depth, $h$, and the corresponding critical depth, $h_c$ \cite{Miles1976}.

Of particular interest is the evolution of weakly nonlinear waves steadily propagating around a circular cylinder of unit radius, focusing on the case where the fluid depth is precisely equal to the critical depth of a 1:2 resonance \cite{Yang2021}.
For the complex-valued eigenmodes defined in equation \eqref{eq:Bessel_eig}, the correlation condition, $\iint_{\mathcal{D}} \Psi_1^2 \Psi_2^* \dA \neq 0$, determines that the angular wavenumbers satisfy $m_2 = 2m_1$ \cite{ChossatDias1995, Yang2021}. By expressing the complex wave amplitudes in polar form, $A_j(\tau) = a_j(\tau) \me^{\mi \theta_j(\tau)}$ (for $j = 1,2$), equation \eqref{eq:Wilton_ripples_amplitude_eqs} may be recast as \cite{Miles1976}
\[ \sd{a_1}{\tau} = -\gamma a_1 a_2 \cos\Theta, \quad 
\sd{a_2}{\tau} = \frac{\gamma}{4}a_1^2 \cos\Theta, \quad \sd{\Theta}{\tau} =  2\gamma a_2 \bigg[1 - \frac{a_1^2}{8a_2^2}\bigg] \sin\Theta, \]
where $\Theta(\tau) = \theta_2(\tau) - 2\theta_1(\tau)$ is the time-dependent phase shift. Steadily propagating waves correspond to time-independent solutions for $a_1$, $a_2$ (both nonzero) and $\Theta$, from which we deduce that $\cos\Theta = 0$ and $a_1/a_2 = 2\sqrt{2}$. Indeed, it is remarkable that the amplitude ratio of the two dominant (normalised) wave modes is independent of the angular wavenumbers, $m_j$, the radial wavenumbers, $K_j$, and the corresponding angular frequencies, $\Omega_j$ (see \S \ref{sec:circular_cylinder} for details). Furthermore, one may readily determine the relationship between the angular velocity of the steady wave rotation and the corresponding wave amplitude, which may then be compared to the numerical solution of the full Euler equations \cite{Yang2021}. This comparison, as well as a comparison to steadily propagating waves computed from various truncations of the Euler equations, will be the subject of future investigation.

\section{The excitation of resonant triads}
\label{sec:excitation}

Having established the existence and evolution of resonant triads, we now focus on the excitation of a particular triad via external forcing. So as to motivate the method of excitation, we first recall (\S \ref{sec:pump_modes}) the well-known result that one mode in the triad may, or may not, excite the other two modes \cite{Davis1967, Hasselmann1967, Simmons1969}; in the case of excitation, the initial mode is referred to as the \emph{pump} mode \cite{CraikBook}. We will then utilise the criterion of the pump mode to excite all three modes in the triad via a pulsating pressure source (\S \ref{sec:pressure_source}). Throughout this section, we continue with the convention that the triad angular frequencies satisfy $\Omega_1 + \Omega_2 + \Omega_3 = 0$, as set forth in \S \ref{sec:triad_eqs}.


\subsection{Excitation via the triad pump mode}
\label{sec:pump_modes}

To first identify the triad pump mode and then characterise the resultant excitation, we consider the case for which $A_3$, say, is much larger in magnitude than the other two mode amplitudes, so $|A_1|, |A_2| \ll |A_3|$ \cite{Davis1967, Hasselmann1967, Simmons1969}. By linearising the triad equations \eqref{eq:triad_equations}, we obtain
\begin{equation}
\label{eq:pump_approx}
\sd{A_1}{\tau} = \alpha_1 A_2^* A_3^*, \qquad 
\sd{A_2}{\tau} = \alpha_2 A_1^* A_3^*, \qquad 
\sd{A_3}{\tau} = 0,
\end{equation}
from which we immediately conclude that $A_3$ is constant (whilst the linearisation assumption holds); we denote $A_3(\tau) = C$ for some given complex number $C$. By considering second derivatives of $A_1$ and $A_2$, we deduce the linearised evolution equations \cite{CraikBook}
\[ \sd{{}^2A_1}{\tau^2} = \alpha_1\alpha_2 |C|^2 A_1 
\quad\mathrm{and}\quad 
\sd{{}^2A_2}{\tau^2} = \alpha_1\alpha_2 |C|^2 A_2,
\]
where $\alpha_1\alpha_2 = \mathscr{C}^2\beta^2/(4\Omega_1\Omega_2)$ (see equation \eqref{eq:triad_coeff_final}). We conclude that $A_1(\tau)$ and $A_2(\tau)$ grow exponentially in time (whilst the linearisation approximation holds) when $\Omega_1 \Omega_2 > 0$, and exhibit sinusoidal oscillations when $\Omega_1\Omega_2 < 0$ \cite{Davis1967, Hasselmann1967, CraikBook}. Thus, mode 3 may excite modes 1 and 2 when $\Omega_1$ and $\Omega_2$ have the same sign (and likewise for other mode permutations). As one angular frequency must have a different sign from the other two (so as to satisfy $\Omega_1 + \Omega_2 + \Omega_3 = 0$), we conclude that the mode whose angular frequency is largest in magnitude (i.e.\ differs in sign) is the triad pump mode \cite{CraikBook}. Equivalently, the pump mode is the mode with largest wavenumber, $K_j$, providing a robust mechanism for an inverse energy cascade to lower wavenumbers \cite{Annenkov2006}.
\begin{figure}
\centering
\includegraphics[width=1\textwidth]{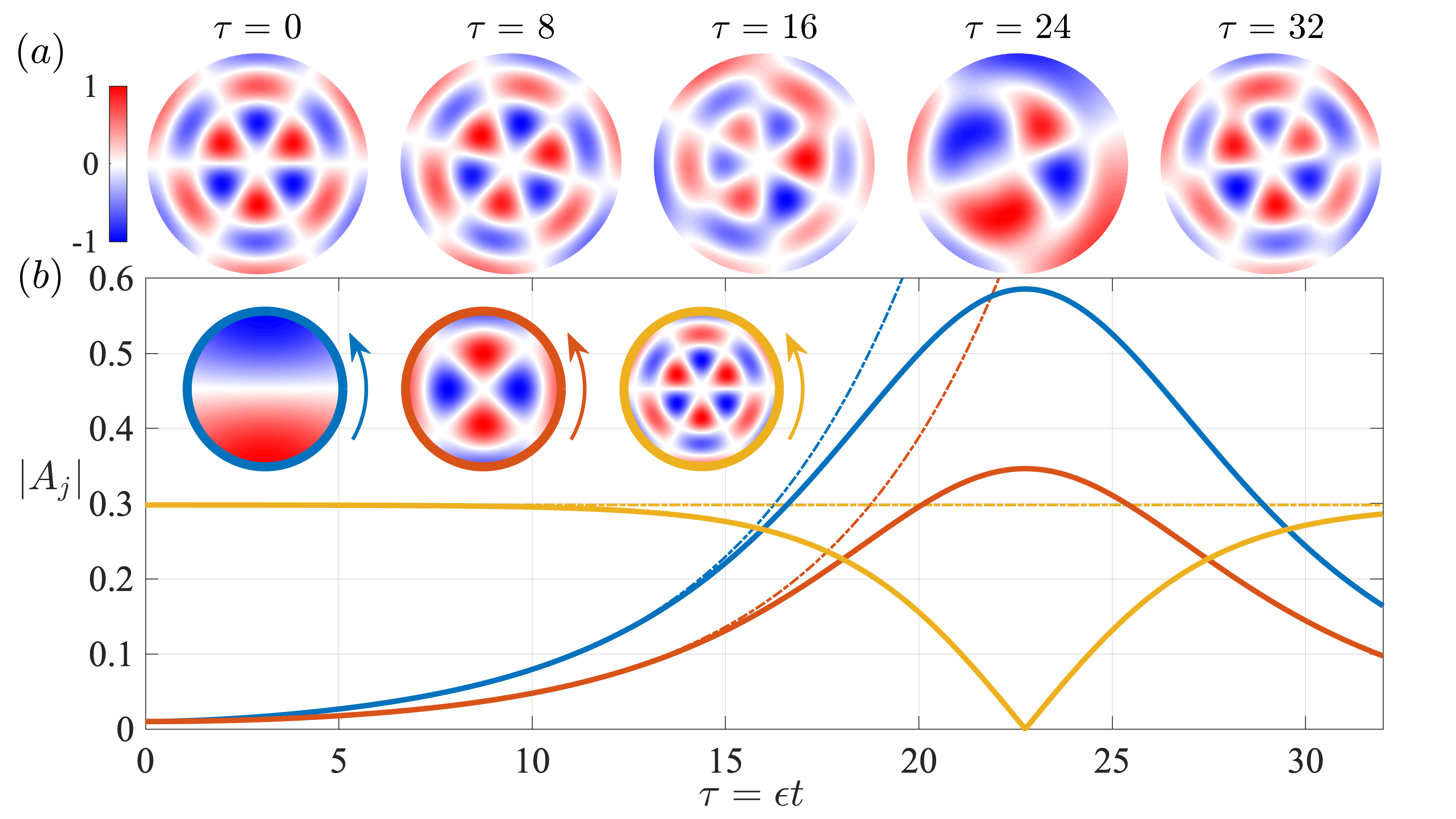}
\caption{\label{fig:Triad_pump_visualisation}
Excitation of a triad via its pump mode for the case of a circular cylinder. We consider \refone{triad 11} in table \ref{Table_circle}, but with $m_3 \mapsto -m_3$.
We choose $\Omega_1, \Omega_2 > 0$ and $\Omega_3 < 0$, so that mode 3 is the pump mode.
$(a)$ Evolution of the free-surface, $\eta \sim -u_t$, over the slow time-scale, $\tau = \epsilon t$, with $\epsilon  = 10^{-3}$. 
$(b)$ The evolution of the wave amplitudes, $|A_j|$, according to the triad equations (equation \eqref{eq:triad_equations}, solid curves) and the pump-mode approximation (equation \eqref{eq:pump_approx}, dashed-dotted curves). Insets: modes 1 (blue),  2 (red) and 3 (gold) at $\tau = 0$; all three modes rotate counter-clockwise.
The simulations were initialised from $A_1(0) = 0.01$ and $A_2(0) = 0.01\mi$, where $A_3(0)$ was chosen to be the positive real number satisfying $\mathscr{E}_1 + \mathscr{E}_2 + \mathscr{E}_3 = 1$, with $\mathscr{E}_j = \Omega_j^2 |A_j|^2$ (see \S \ref{sec:simplify_coeff}).
}
\end{figure}

To visualise the influence of the pump mode on the resultant free-surface pattern, we present the solution of the triad equations \eqref{eq:triad_equations} and the corresponding pump-mode approximation (equation \eqref{eq:pump_approx}) in figure \ref{fig:Triad_pump_visualisation}. By recalling that the free surface satisfies $\eta = -u_t + O(\epsilon)$, we first deduce that
\begin{equation}
\label{eq:eta_exp}
\eta(\x,t) \sim \sum_{j = 1}^3 \Big[\mi \Omega_j A_j(\tau) \Psi_j(\x) \me^{-\mi \Omega_j t} + \cc\Big] + O(\epsilon).
\end{equation}
For the case of a circular cylinder, we utilise the complex-valued eigenmodes defined in equation \eqref{eq:Bessel_eig}, corresponding to the superposition of steadily propagating waves for $m_j \neq 0$ (the rotation direction depends on the sign of $\Omega_j /m_j$). Upon initialising the system so that the energy is primarily within the pump mode (mode 3), modes 1 and 2 are gradually excited due to nonlinear interaction, with exponential growth evident for $\tau \lesssim 10$. As time further increases, the dynamics depart from the pump-mode approximation: the energy in the pump mode appreciably decreases, whilst the energy in modes 1 and 2 saturates. The free surface varies qualitatively during this evolution, with an appreciable change in pattern structure visible by $\tau = 24$ (primarily a superposition of modes 1 and 2). Notably, the system evolution is periodic, which becomes apparent over longer time scales.

\subsection{Excitation via an applied pressure source}
\label{sec:pressure_source}

Based on the ideas of the previous section, we consider a methodology for exciting the pump mode of a triad, which will subsequently excite the remaining two modes (provided that the initial disturbance of each of the remaining modes is nonzero).
Notably, several methods for exciting internal resonances have been considered in prior investigations, primarily focusing on imposed motion of the fluid vessel via horizontal \cite{Miles1976, Miles1984c} or vertical vibration \cite{Miles1976, Miles1984b, MilesHenderson1990, HendersonMiles1991}. Furthermore, one may, in principle, utilise sinusoidal paddles or plungers to excite a particular triad's pump mode for a given geometry (similar wave makers are used in rectangular wave tanks \cite{McGoldrick1970b, HendersonHammack1987}). However, for large-scale fluid tanks, imposed motion of the vessel may be impractical (if the tank were set in a concrete base, for example), and it may be challenging to determine the correct paddle motion necessary to excite a chosen pump mode for geometrically complex cylinders. We choose, therefore, to consider a slightly different approach: we instead excite the pump mode via a pulsating pressure source located just above the free surface (e.g.\ an air blower). 

In order to incorporate a pressure source within our mathematical framework, we first reformulate the dimensionless dynamic boundary condition (equation \eqref{eq:Euler_DBC}) as
\[ \phi_t + \eta + \frac{\epsilon}{2}\Big(|\nabla \phi|^2 + \phi_z^2\Big)  + \epsilon P(\x, t) = 0 \quad \mbox{for}\quad \x \in \mathcal{D}, \quad z = \epsilon \eta, \]
where the dimensional pressure is $\epsilon^2 a \rho g P$ for fluid density $\rho$ ($P = 0$ corresponds to atmospheric pressure). The pressure source is chosen to be small in magnitude so that the resultant wave excitation arises over the slow time-scale, $\tau = \epsilon t$, and may thus be saturated by weakly nonlinear effects. By modifying the developments outlined in \S \ref{sec:BL_eq}, we derive the forced Benney-Luke equation
\begin{equation}
\label{eq:BL_eq_pressure}
u_{tt} + \DtN u + \epsilon\bigg(
u_t\big(\DtN^2 + \Delta\big)u + \pd{}{t}\Big[(\DtN u)^2 + |\nabla u|^2\Big] + \partial_t P\bigg) =  O(\epsilon^2) \quad\mathrm{for}\quad \x \in \mathcal{D},
\end{equation}
which will be the starting point for the asymptotic analysis.

Before proceeding further, we first describe two forms of the pressure source relevant to our investigation.
For a stationary pressure source oscillating periodically over the fast time-scale, $t$, we express $P(\x,t) = f(\tau) s(\x) \me^{-\mi \Omega_p t} + \cc$, where $s(\x)$ is a fixed spatial profile (generally spanning the cavity), $f(\tau)$ accounts for a slow modulation in the magnitude of the pressure, and $\Omega_p$ is the pulsation angular frequency. We choose $\Omega_p$ to be close to the angular frequency of the pump mode, which, without loss of generality, we assume to be mode 3 (i.e.\ $\Omega_3$ has the opposite sign from $\Omega_1$ and $\Omega_2$). We denote, therefore, $\Omega_p = \Omega_3 + \epsilon \mu$, where $\mu = O(1)$ determines the extent of the frequency mismatch. 
For a pressure source orbiting the centre of a circular cylinder at a constant angular velocity, we instead posit that $P$ has the form $P(r,\theta, t) = f(\tau) s(r, \theta - \Omega_p t)$, where $\Omega_p = (\Omega_3 + \epsilon \mu)/m_3$ is the angular velocity of the pressure source (assuming that the pump mode is non-axisymmetric, i.e.\ $m_3 \neq 0$).

For both standing and orbiting pressure sources, we now follow a similar multiple-scales procedure to that outlined in \S \ref{sec:multiple_scales}, starting from the forced Benney-Luke equation \eqref{eq:BL_eq_pressure}.
So as to discount the possibility that the pressure source excites more than one mode in the triad, we assume that neither $|\Omega_1|$ or $|\Omega_2|$ are close to $|\Omega_3|$. Furthermore, we incorporate a weak detuning in the triad angular frequencies, denoting $\Omega_1 + \Omega_2 + \Omega_3 = \epsilon \sigma$ (see \S \ref{sec:weak_detuning}). 
It follows that each complex amplitude, $A_j(\tau)$, evolves according to
\begin{equation}
\label{eq:triad_forced}
\sd{A_1}{\tau} = \alpha_1 A_2^* A_3^* \me^{\mi \sigma \tau}, \quad
\sd{A_2}{\tau} = \alpha_2 A_1^* A_3^* \me^{\mi \sigma \tau}, \quad
\sd{A_3}{\tau} = \alpha_3 A_1^* A_2^* \me^{\mi \sigma \tau} - \Omega_3s_3f(\tau)\me^{-\mi \mu \tau},
\end{equation}
where the coefficients, $\alpha_j$, are defined in equation \eqref{eq:triad_coeff_final}. Notably, the pump mode may only be excited provided that the corresponding eigenmode is non-orthogonal to the pressure source, corresponding to a nonzero projection, i.e.\ $s_3\neq 0$, where $s_3 = \langle \Psi_3, s\rangle$. Similar equations describing the evolution of forced resonant triads have been explored by McEwan \emph{et al.}\ \cite{McEwan1972} (with the inclusion of linear damping) and Raupp \& Silva Dias \cite{Raupp2009}.

\begin{figure}
\centering
\includegraphics[width=1\textwidth]{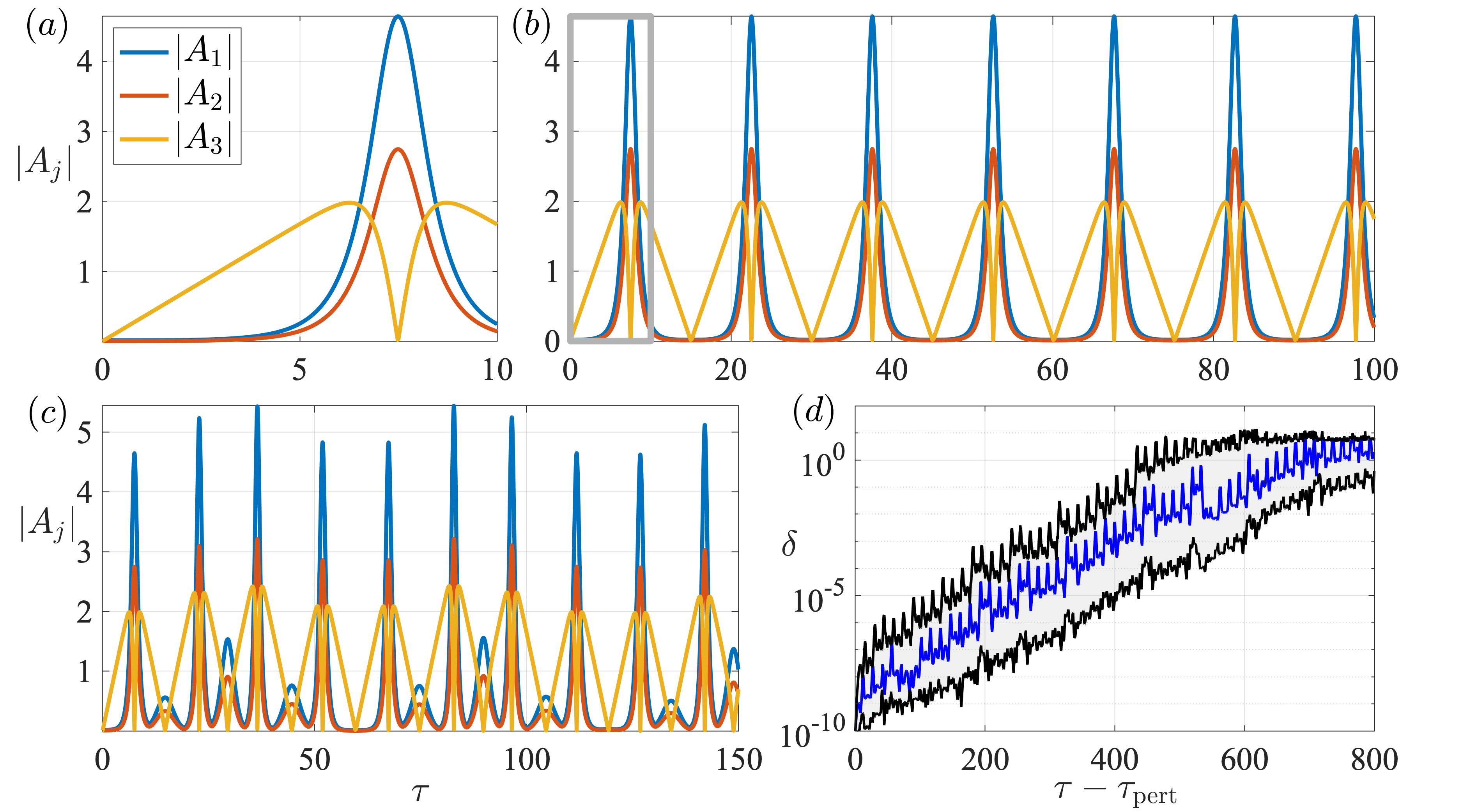}
\caption{\label{fig:Forced_triad}
Evolution of the forced triad equations \eqref{eq:triad_forced} for $\sigma = \mu = 0$ and constant $f$. We consider the same triad as figure \ref{fig:Triad_pump_visualisation}, with $s_3 f = 0.1$. In all three panels, $A_1(0) = 0.02\mi$ and $A_2(0) = 0.01$.
For $A_3(0) = 0.01$, we observe $(a)$ the initial excitation of the triad and $(b)$ the resultant periodic dynamics (the initial growth is highlighted within the grey box).
$(c)$ For $A_3(0) = 0.01\mi$, the triad evolution appears chaotic.
\reftwo{$(d)$ The separation distance, $\delta(\tau)$, of a trajectory randomly perturbed at time $\tau_{\mathrm{pert}}$, with $\delta(\tau_{\mathrm{pert}}) = 10^{-10}$ and the same initialisation (at $\tau = 0$) as $(c)$. The black curves bound $\delta(\tau)$ for 30 equally spaced values of $\tau_{\mathrm{pert}}$ in the interval $100 \leq \tau_{\mathrm{pert}} \leq 115$; the blue curve corresponds to $\tau_{\mathrm{pert}} = 100$.}
}
\end{figure}

In the special case of time-independent forcing ($f$ constant) and no frequency detuning ($\sigma = \mu = 0$), the dynamics of the forced triad equations has been analysed by Harris \emph{et al.}\ \cite{Harris2012}, with both periodic and quasi-periodic dynamics reported. We also consider this case, leaving the effects of detuning and variable forcing for future investigation. In this setting, when $|A_1|$, $|A_2|$ and $|A_3|$ are initially small relative to the magnitude of the forcing, $|\Omega_3 s_3 f|$, the initial growth in $A_3$ is approximately linear (see figure \ref{fig:Forced_triad}$(a)$). As mode 3 is the pump mode, the growth in $A_3$ excites $A_1$ and $A_2$, thus activating the triad. The conservation laws of the forced triad equations \cite{Harris2012} result in a temporary diminution of mode 3, which is later augmented by the external forcing; whence the process repeats. In some parameter regimes, the resulting evolution of the forced triad is periodic in time (see figure \ref{fig:Forced_triad}$(b)$ and Raupp \& Silva Dias \cite{Raupp2009}); in contrast to the findings of Harris \emph{et al.}\ \cite{Harris2012}, however, we also identify initial conditions (with all other parameters unchanged) that result in hitherto unidentified chaotic dynamics (see figure \ref{fig:Forced_triad}$(c)$). 

\reftwo{
To verify the chaotic nature of this latter example, we consider the separation distance of two initially adjacent trajectories in phase space, for which we observe exponential divergence in time (see figure \ref{fig:Forced_triad}$(d)$). This exponential divergence is indicative of a positive maximal Lyapunov exponent \cite{Strogatz}, which characterises the sensitivity to initial conditions exhibited by chaotic systems. Specifically, for a solution, $A_j(\tau)$, and its perturbation, $A_{j,\mathrm{pert}}(\tau)$, we consider the evolution of the separation distance, defined
\[ 
\delta(\tau) = \sqrt{\sum_{j = 1}^3 \big|A_j(\tau) - A_{j,\mathrm{pert}}(\tau)\big|^2 }.
\] 
To compute $\delta(\tau)$, we first simulate the forced triad equations \eqref{eq:triad_forced} on the interval $0 \leq \tau \leq \tau_{\mathrm{pert}}$, 
with $A_j(0) = A_{j,\mathrm{pert}}(0)$ and $\tau_{\mathrm{pert}}$ chosen to be sufficiently large so as to ensure that the chaotic attractor (should one exist) be approached. At $\tau = \tau_{\mathrm{pert}}$, the real and imaginary parts of $A_{j, \mathrm{pert}}$ are both randomly perturbed according to a uniform distribution on the interval $(-1, 1)$, with the resulting complex perturbation scaled so that $\delta(\tau_{\mathrm{pert}}) = 10^{-10}$. We then evolve $A_j(\tau)$ and $A_{j,\mathrm{pert}}(\tau)$ for $\tau \geq \tau_{\mathrm{pert}}$, giving rise to initial exponential growth of $\delta(\tau)$, with saturation when the perturbation distance is comparable to the `diameter' of the chaotic attractor (see figure \ref{fig:Forced_triad}$(d)$).}

\reftwo{
To confirm that the exponential growth was not specific to a perturbation about a particular point on the chaotic attractor \cite{Strogatz}, we considered 30 equally spaced values of $\tau_{\mathrm{pert}}$ on the interval $100 \leq \tau_{\mathrm{pert}} \leq 115$, roughly corresponding to the time taken for one `loop' of the chaotic attractor to take place (see figure \ref{fig:Forced_triad}$(c)$). Each simulation was computed with a fourth-order Runge-Kutta method and a time step of 0.005.
For each value of $\tau_{\mathrm{pert}}$, we observed similar exponential divergence of trajectories (see figure \ref{fig:Forced_triad}$(d)$); moreover, the evolution of $\delta(t)$ during the growth phase remained unchanged when the time step was decreased to 0.001, with numerical errors only accumulating over longer time scales.
Our results thus provide strong evidence that there is a positive maximal Lyapunov exponent in this particular portion of parameter space, indicative of chaotic dynamics.}

\section{Discussion}
\label{sec:discussion}

We have performed a systematic investigation into nonlinear resonant triads of free-surface gravity waves confined to a cylinder of finite depth; previously studied 1:2 resonances are obtained as special cases. A key result of our study is Theorem \ref{thm:triads}, which determines whether there exists a fluid depth at which three given wave modes resonate due to the nonlinear evolution of the fluid. Equipped with this result, we determined the long-time fluid evolution using multiple-scales analysis, from which we deduced that all solutions to the triad equations are periodic in time. Finally, we determined that a given triad may be excited via external forcing of the triad's pump mode, thereby providing a mechanism for exciting a given triad in a wave tank. All our results are derived for cylinders of arbitrary cross-section (barring some technical assumptions; see \S \ref{sec:formulation}), thus forming a broad framework for characterising nonlinear resonance of confined free-surface gravity waves. In particular, our theoretical developments buttress experimental observations \cite{Michel2019} and demonstrate the potential generality of confinement as a mechanism for promoting nonlinear resonance.

A second fundamental component of our study is the influence of the cylinder cross-section on the existence of resonant triads; for example, resonant triads are impossible in rectangular cylinders, yet abundant within circular and annular cylinders (for particular fluid depths). Of the vast array of resonances arising in a circular cylinder, triads consisting of an axisymmetric pump mode and two identical counter-propagating waves are of notable interest. This combination of axisymmetric and non-axisymmetric modes possesses an interesting analogy to the excitation of counter-propagating subharmonic beach edge waves due to a normally incident standing wave \cite{Guza1974}. Specifically, the wave crests of the standing axisymmetric mode are always parallel to the bounding wall of the circular cylinder, and may excite steadily propagating waves that are periodic in the azimuthal direction. For the special case for which the amplitudes of the two counter-propagating modes coincide, one observes the resonant interaction of standing axisymmetric and non-axisymmetric waves.

So as to gain a deeper insight into the influence of nonlinearity on resonant triads, a primary focus for future investigations will be the simulation of the Euler equations within a cylindrical domain, with consideration of various truncated systems \cite{CraigSulem1993, MilewskiKeller1996, BergerMilewski2003, WangMilewski2012}. From a computational perspective, the most natural geometry to consider is a circular cylinder \cite{QadeerWilkening2019}; this geometry has been previously explored in the context of steadily propagating nonlinear waves in the vicinity of a 1:2 resonance \cite{Bryant1989, Yang2021}, but it remains to assess the efficacy of the amplitude equations \eqref{eq:triad_equations} for predicting the evolution of nonlinear triads. Indeed, exploration of the nonlinear dynamics may reveal additional resonant triads arising beyond the small-wave-amplitude limit explored herein.
Of similar interest is the fluid evolution when multiple triads are excited at a single depth, with the potential for energy exchange via triad-triad interactions \cite{McEwan1972, CraikBook, Chow1996, Choi2021}. 
The simulation of free-surface gravity waves in non-circular cylinders presents a more formidable challenge, however, except for cylinder cross-sections that possess a tractable eigenmode decomposition.

A second natural avenue for future investigation is to characterise the influence of applied forcing on resonant triads. For example, when the fluid bath is subjected to sufficiently vigorous vertical vibration, Faraday waves \cite{Faraday1831, Kumar1996} may appear on the free surface; although this scenario has been studied in the case of a 1:2 internal resonance \cite{Miles1984b, MilesHenderson1990, HendersonMiles1991}, resonant triads may give rise to the formation of more exotic free-surface patterns, particularly at fluid depths that differ from that of a 1:2 resonance. In a similar vein, horizontal vibration \cite{Miles1976, Miles1984c} or a pulsating pressure source at the frequency of the triad's pump mode may lead to a wealth of periodic and quasi-periodic dynamics, as predicted by the forced triad equations \cite{Harris2012}. Our study has indicated, however, that chaotic dynamics are also possible in some parameter regimes, and might thus be excited in numerical simulation or experiments. Lastly, our study has focused on flat-bottomed cylinders; it seems plausible, however, that submerged topography may enhance or mitigate certain resonances, which may be an important consideration in the design of industrial-scale fluid tanks.

Finally, our study has focused on the special case of a liquid-air interface, for which the dynamics of the air are neglected within the Euler equations. It is natural, however, to extend our formulation to the case of two-layer flows (in the absence of surface tension), with two immiscible fluids (e.g.\ air and water) confined within a cylinder whose lid and base are both rigid. In this setting, the density difference across the fluid-fluid interface has a strong influence of the system dynamics; it seems plausible, therefore, that additional resonances may be excited in this configuration, relative to the liquid-air interface considered herein. 
Notably, the anticipated resonances would arise across a single interface, in contrast to the cross-interface resonances explored in previous investigations \cite{Ball1964, Simmons1969, Joyce1974, Segur1980, TakloChoi2019, Choi2021}. Finally, exploring the influence of parametric forcing \cite{KumarTuckerman1994} on resonant triads arising for two-layer flows opens up exciting new vistas in nonlinear resonance induced by confinement.

\appendix

\section{Proof of Theorem \ref{thm:triads}}
\label{app:thm_proof}
\begin{proof}
To prove Theorem \ref{thm:triads}, we first show that there are no values of $h \in (0, \infty)$ satisfying $\Omega_1 + \Omega_2 = \Omega_3$ when $K_1 + K_2 \geq K_3$ or when $\sqrt{K_1} + \sqrt{K_2} \leq \sqrt{K_3}$, where we recall that $\Omega_j(h) = \sqrt{K_j \tanh(K_j h)}$ and $K_j > 0$ for $j = 1,2,3$. We then prove that there exists a solution to $\Omega_1 + \Omega_2 = \Omega_3$ when $K_1 + K_2 < K_3 < (\sqrt{K_1} + \sqrt{K_2})^2$, and that this solution is unique.

In the case $K_1 + K_2 \geq K_3$, we first define $\chi(K;h) = \sqrt{K\tanh(Kh)}$. For fixed $h > 0$, we observe that 
\[\chi(K_3;h) \leq \chi(K_1 + K_2;h) < \chi(K_1;h) + \chi(K_2;h), \]
where we have utilised that $\chi(K;h)$ is a positive, monotonically increasing, concave function of $K > 0$. We conclude that $\Omega_3 < \Omega_1 + \Omega_2$ for any $h > 0$, so there are no values of $h$ for which $\Omega_1 + \Omega_2 = \Omega_3$.

In the case $\sqrt{K_1} + \sqrt{K_2} \leq \sqrt{K_3}$, we first note that the lower bound $K_j > 0$ (for $j = 1,2,3$) implies that $K_1 < K_3$ and $K_2 < K_3$. Furthermore, as $\tanh(x)$ is a monotonically increasing function for $x > 0$, we conclude that $\tanh(K_j h) < \tanh(K_3 h)$ for $j = 1,2$ and all $h > 0$. We now utilise this property to deduce that
\[ \sqrt{K_1 \tanh(K_1 h)} + \sqrt{K_2 \tanh(K_2 h)} < \big(\sqrt{K_1} + \sqrt{K_2}\big)\sqrt{\tanh(K_3h)} \leq \sqrt{K_3 \tanh(K_3 h)}. \]
We conclude that $\Omega_3 > \Omega_1 + \Omega_2$ for any $h > 0$, so there are no values of $h$ for which $\Omega_1 + \Omega_2 = \Omega_3$.

For the remainder of the proof, we consider the case 
\begin{equation}
\label{app:kineq}
K_1 + K_2 < K_3 \quad\mathrm{and}\quad \sqrt{K_3} < \sqrt{K_1} + \sqrt{K_2},
\end{equation}
which is equivalent to the pair of inequalities given by equation \eqref{eq:kineq}.
Indeed, we will show that there exists a unique value of $h > 0$ satisfying $\Omega_1 + \Omega_2 = \Omega_3$ in this case. Equivalently, we demonstrate that $F(h) = \big(\Omega_1(h) + \Omega_2(h)\big)/\Omega_3(h) -1$ has a unique positive root, where we express
\[ F(h) = \sqrt{\psi_1(h)} + \sqrt{\psi_2(h)} - 1, \]
with the positive functions $\psi_1$ and $\psi_2$ defined
\[ \psi_j(h) = \frac{K_j \tanh(K_j h)}{K_3\tanh(K_3 h)} \quad\mathrm{for}\,\,\, j = 1,2. \]

In order to show the existence of a root of $F(h)$, we first note that
\[\lim_{h \rightarrow 0} F(h) = \frac{K_1 + K_2}{K_3} - 1 < 0 
\quad\mathrm{and}\quad 
\lim_{h \rightarrow \infty} F(h) = \frac{\sqrt{K_1} + \sqrt{K_2}}{\sqrt{K_3}} - 1 > 0, \]
where we have used the limits $\lim_{x\rightarrow 0} (\tanh(x)/x) = 1$ and $\lim_{x\rightarrow \infty } \tanh(x) = 1$, respectively, and implemented the inequalities given in equation \eqref{app:kineq}. As $F(h)$ is a continuous function, the intermediate-value theorem determines that $F(h)$ has at least one positive root.

To prove that such a root is unique, we demonstrate that $F(h)$ is a strictly monotonically increasing function for $h > 0$. Specifically, we note that (for $j = 1,2$)
\[\sd{\psi_j}{h} = 2 K_3 \psi_j(h)\bigg(\frac{K_j}{K_3}\mathrm{cosech}(2K_j h) - \mathrm{cosech}(2K_3 h)\bigg) > 0 \quad \mathrm{for}\,\,\, 0 < K_j < K_3, \]
where the inequality follows from the convexity of $\mathrm{cosech}(x)$ for $x > 0$, i.e.\ $b\, \mathrm{cosech}(bx) > \mathrm{cosech}(x)$ for $0 < b < 1$ and all $x > 0$ (associating $x = 2 K_3 h$ and $b = K_j / K_3$). 
As the bounds $K_1 < K_3$ and $K_2 < K_3$ incorporate the region determined by equation \eqref{app:kineq}, we deduce that $F(h)$ is strictly monotonically increasing. We conclude, therefore, that the root of $F(h)$ must be unique, thereby completing the proof.

\end{proof}

\section{Wavenumbers in an annulus}
\label{app:annulus_wavenumbers}

The no-flux condition (equation \eqref{eq:Euler_no_flux_walls}) on the inner and outer radii of an annulus requires that $\partial_r \Phi_{mn}(r_0,\theta) = 0 $ and $\partial_r \Phi_{mn}(1,\theta) = 0$ for all $\theta$, where $\Phi_{mn}(r,\theta)$ is the cylinder function defined in equation \eqref{eq:cylinder_functions}. It follows, therefore, that the corresponding wavenumber, $k_{mn}$, and weighting factor, $\gamma_{mn}$, satisfy the equations
\begin{subequations}
\label{eq:annulus_no_flux_app}
\begin{align}
\Jb_m'(k_{mn}r_0) \cos(\gamma_{mn}\pi) + \Yb_m'(k_{mn}r_0)\sin(\gamma_{mn}\pi) &= 0, \label{eq:annulus_no_flux_r0} \\
\Jb_m'(k_{mn}) \cos(\gamma_{mn}\pi) + \Yb_m'(k_{mn})\sin(\gamma_{mn}\pi) &= 0. \label{eq:annulus_no_flux_1}
\end{align}
\end{subequations}
By rearranging equation \eqref{eq:annulus_no_flux_app}, we determine the following expressions for $\tan(\gamma_{mn}\pi)$:
\begin{equation}
\label{eq:annulus_no_flux_gamma}
\tan(\gamma_{mn}\pi) = -\frac{\Jb'_m(k_{mn}r_0)}{\Yb'_m(k_{mn}r_0) } 
\quad\mathrm{and}\quad
 \tan(\gamma_{mn}\pi) = -\frac{\Jb'_m(k_{mn})}{\Yb'_m(k_{mn}) }.
\end{equation}
By eliminating $\tan(\gamma_{mn}\pi)$ and rearranging, we find that $k_{mn} > 0$ satisfies equation \eqref{eq:annulus_no_flux}. Upon computing $k_{mn}$, one may then determine $\gamma_{mn} \in [0 ,1]$ using either of the equivalent expressions for $\tan(\gamma_{mn} \pi)$ given in equation \eqref{eq:annulus_no_flux_gamma}.

\section{Reduction of the triad coefficients}
\label{app:reduce_triad_coeff}

As motivated by the form of $\alpha_1$ given in equation \eqref{eq:alpha1_simp}, we demonstrate that 
\begin{equation}
\label{app:simp_eq1}
\Omega_2 L_3^2 + \Omega_3L_2^2 - 2\Omega_1 L_2 L_3 = -\frac{1}{2}\Omega_1\Omega_2\Omega_3\big(\Omega_1^2 + \Omega_2^2 + \Omega_3^2\big),
\end{equation}
where we recall that $\Omega_1 + \Omega_2 + \Omega_3 = 0$ and $L_j = \Omega_j^2$.  In fact, the equality given in equation \eqref{app:simp_eq1} holds under cyclic permutation of the indices $(1,2,3)$ (as is necessary when defining $\alpha_2$ and $\alpha_3$), where we note that the right-hand side is unchanged under such permutations. We conclude that $\alpha_2$ and $\alpha_3$ may be simplified in a similar manner, with the right-hand side of equation \eqref{app:simp_eq1} appearing as a constant term in all three coefficients (see \S \ref{sec:simplify_coeff}).

We now detail the algebraic manipulations necessary to transform the left-hand side of equation \eqref{app:simp_eq1} into the right-hand side. By substituting $L_j = \Omega_j^2$ into the left-hand side of equation \eqref{app:simp_eq1} and factorising, we obtain
\begin{equation}
\label{app:simp_eq2}
\Omega_2 L_3^2 + \Omega_3L_2^2 - 2\Omega_1 L_2 L_3 = \Omega_2^4\Omega_3 + \Omega_2\Omega_3^2\big(\Omega_3^2 - 2\Omega_1\Omega_2\big).
\end{equation}
Next, we substitute
\begin{equation}
\label{app:Omega_3_sq}
\Omega_3^2 = (\Omega_1^2 + \Omega_2^2) = \Omega_1^2 + 2\Omega_1\Omega_2 + \Omega_2^2
\end{equation}
into equation \eqref{app:simp_eq2}, yielding
\begin{equation}
\label{app:simp_eq3}
\Omega_2 L_3^2 + \Omega_3L_2^2 - 2\Omega_1 L_2 L_3 = \Omega_2\Omega_3\Big[\Omega_2^3 + \Omega_3\big(\Omega_1^2 + \Omega_2^2\big)\Big].
\end{equation}
We proceed by substituting $\Omega_3 = -(\Omega_1 + \Omega_2)$ within the square brackets in equation \eqref{app:simp_eq3}; by distributing and cancelling common terms, we obtain 
\begin{equation}
\label{app:simp_eq4}
\Omega_2 L_3^2 + \Omega_3L_2^2 - 2\Omega_1 L_2 L_3 = -\Omega_1\Omega_2\Omega_3\Big[\Omega_1^2 + \Omega_1\Omega_2 + \Omega_2^2\Big].
\end{equation}
Finally, we rearrange equation \eqref{app:Omega_3_sq} to give
\[ \Omega_1\Omega_2 = \frac{1}{2}\Big(\Omega_3^2 - \Omega_1^2 - \Omega_2^2\Big), \]
which, upon substitution into equation \eqref{app:simp_eq4}, supplies the required result (equation \eqref{app:simp_eq1}).

\bibliography{TriadsBib}
\bibliographystyle{plain}

\end{document}